\documentclass[apj]{emulateapj}
\usepackage{natbib}
\usepackage{amsmath}
\usepackage{apjfonts}

\newcommand{\dev}{\mathrm{d}}

\begin{document}

\title{Gamma-ray burst afterglow broadband fitting based directly on hydrodynamics simulations}

\author{Hendrik van Eerten$^1$, Alexander van der Horst$^2$, Andrew MacFadyen$^1$}
\affil{
  $^1$ Center for Cosmology and Particle Physics, Physics Department, New York University, New York, NY 10003\\
  $^2$ Universities Space Research Association, NSSTC, 320 Sparkman Drive, Huntsville, AL 35805, USA
}

\begin{abstract}
We present a powerful new tool for fitting broadband gamma-ray burst afterglow data, which can be used to determine the burst explosion parameters and the synchrotron radiation parameters. By making use of scale invariance between relativistic jets of different energies and different circumburst medium densities, and by capturing the output of high-resolution two-dimensional relativistic hydrodynamical (RHD) jet simulations in a concise summary, the jet dynamics are generated quickly. Our method calculates the full light curves and spectra using linear radiative transfer sufficiently fast to allow for a direct iterative fit of RHD simulations to the data. The fit properly accounts for jet features that so far have not been successfully modeled analytically, such as jet decollimation, inhomogeneity along the shock front and the transitory phase between the early time relativistic and late time non-relativistic outflow. As a first application of the model we simultaneously fit the radio, X-ray and optical data of GRB 990510. We not only find noticeable differences between our findings for the explosion and radiation parameters and those of earlier authors, but also an improved model fit when we include the observer angle in the data fit. The fit method will be made freely available on request and on-line at \url{http://cosmo.nyu.edu/afterglowlibrary}. In addition to data fitting, the software tools can also be used to quickly generate a light curve or spectrum for arbitrary observer position, jet and radiation parameters.
\end{abstract}


\section{Introduction}
\label{introduction_section}
Gamma-ray bursts (GRBs) are short intense flashes of gamma radiation produced by cataclysmic stellar events such as the collapse of the core of a massive star (\citealt{Woosley1993, Paczynski1998, MacFadyen1999}) or a neutron star-neutron star or neutron star-black hole merger (e.g. \citealt{Eichler1989, Paczynski1991}). During these events a collimated relativistic outflow is produced which sweeps up the matter surrounding the GRB. Regardless of the original mass content or launching mechanism of this outflow (be it a fireball, \citealt{Meszaros1997}, or a Poynting flux dominated jet, e.g. \citealt{Drenkhahn2002}), the expanding blast wave will sweep up circumburst matter and will eventually start to decelerate. As the blast wave shocks the circumburst medium, broadband synchrotron radiation is produced by shock-accelerated electrons, giving rise to an afterglow signal that can be observed for up to days at X-ray and optical frequencies, and for up to years at radio frequencies.

Three kinds of parameters determine the shape of the observed afterglow light curves. First, the shock dynamics are set by the explosion energy and circumburst density. A second set of parameters captures the physics of synchrotron emission from shock-accelerated electrons. Finally, the observed flux depends on the parameters defined by a given observation: frequency, time and observer angle.

Ever since the first afterglows were discovered \citep{Costa1997, Groot1997}, models based on synchrotron radiation from a decelerating relativistic blast wave have been successful in describing the broadband data \citep[e.g.][]{Wijers1997, Sari1998, Wijers1999, Granot2002}. The synchrotron spectrum is typically described as consisting of a number of connected power law regimes, with the critical frequencies connecting the regimes shifting over time and being determined by the basic  spectral shape of synchrotron emission, synchrotron self-absorption and cooling of the shock-accelerated electrons. In order to accurately model the late time afterglow emission in the radio, afterglow models based on a non-relativistic rather than relativistic blast wave have been applied as well \citep[e.g.][]{Frail2000}. Both the early time ultra-relativistic and late time non-relativistic stages of the evolution of the shock are self-similar, with solutions given by \cite{Blandford1976}, hereafter BM, and \cite{Sedov1959}, von Neumann \cite{vonNeumann1961} and \cite{Taylor1950}, hereafter ST, respectively. The intermediate stage can be approximated \citep[e.g.][]{Rhoads1999,Huang1999}, but this stage is complicated by the dynamics of jet decollimation. When jet decollimation is taken into account, a homogeneous jet surface that widens with the comoving speed of sound is often assumed \citep{Rhoads1999}, while more recent jet spreading models \citep{Granot2011} do not take the radial structure of the outflow into account. The jet nature of the outflow ultimately reveals itself as a break in the light curve, the jet break, and a subsequent steepening of the light curve slope. The physics of afterglow jets has been reviewed in e.g. \cite{Piran2005, Meszaros2006, Granot2007}.

Purely analytical models are severely limited in that they do not accurately capture many features of the jet dynamics (such as the aforementioned jet spreading and deceleration) and radiation. The simplifications inherent in a purely analytical approach lead to diverging predictions for a range of features such as the observed shape of the jet break, the size and shape of the counterjet (which was launched away from the observer and is only seen at late times, when relativistic beaming of the emitted radiation plays a lessened role), the nature and duration of the transition to the non-relativistic phase and the effect of the orientation of the jet with respect to the observer. To gain a better understanding of these aspects, various authors have performed numerical relativistic hydrodynamics (RHD) simulations of afterglow jets, in one dimension \citep{Kobayashi1999, Downes2002, Mimica2009, vanEerten2010}, two dimensions \citep{Granot2001, Zhang2006, Meliani2007, Ramirez-Ruiz2010, vanEerten2011, Wygoda2011} and occasionally even three \citep{Cannizzo2004}. Over the past decade, these simulations have steadily increased in accuracy, mainly through the use of adaptive-mesh refinement (AMR) techniques, which locally increase resolution during simulation where needed, and are capable of resolving the six orders of magnitude difference between the initial width of the thin relativistic shell and the late time outer radius of the decelerated jet. Recent high-resolution simulations have shown that relativistic jets spread sideways significantly slower than analytically expected and have a strongly inhomogeneous shock front \citep{Zhang2009, Meliani2010, vanEerten2011, vanEerten2011jetspreading, Wygoda2011}. Furthermore, they show that the transition from relativistic to non-relativistic expansion is a very slow process \citep{Zhang2009, vanEerten2010}; and that jet orientation strongly affects the jet break even for small observer angles \citep{vanEerten2010c}, while for observers both on and off axis the observed jet-break time differs between different frequencies due to synchrotron self-absorption \citep{vanEerten2011}.

A RHD jet simulation can be combined with a numerical synchrotron radiation calculation to yield a powerful tool to predict the evolution of the observable broadband afterglow spectrum in detail. The weaknesses of simplified analytical models are thus avoided and local changes in fluid structure and arrival time effects are correctly accounted for. This calculation can be performed in different ways, for example by summing over the emitted power of all fluid cells in the simulation in the case of an optically thin fluid \citep{Downes2002, Nakar2007bumpyride, Zhang2009} or by fully solving the linear radiative transfer equations including synchrotron self-absorption \citep{vanEerten2010, vanEerten2011sgrbs}.

An obvious drawback of simulation based light curves compared to analytically calculated light curves is that calculating the former is a time consuming process. A full jet simulation takes several thousand CPU-hours to complete. A purely analytical light curve, on the other hand, can be calculated almost instantaneously and can therefore be applied to iterative model fitting, where the procedure of minimizing $\chi^2$ requires at least thousands of light curve calculations with slightly differing explosion and radiation parameters. In this paper, we present a new method to use simulation results directly as a basis for iterative fitting of broadband data, which closes the gap between simulations and analytical models. This method should prove useful for further constraining the physics of gamma-ray burst afterglows, thereby obtaining clues about the nature of the progenitor and the burst environment. This provides more accurate predictions for future surveys including LOFAR \citep{Rottgering2006}, SKA \citep{Carilli2004} or ALMA \citep{Wootten2003}, and indirectly benefits gravitational wave predictions for LIGO \citep{Abbott2009} and VIRGO \citep{Acernese2008}, where GRBs are potentially observable as electro-magnetic counterparts \citep{Nakar2011, vanEerten2011sgrbs}. Finally, our method helps to establish a baseline for studies of the effect of more detailed models of the microphysics (see e.g. \citealt{vanEerten2010, Panaitescu2006, Filgas2011}).

This paper is structured as follows. First we briefly describe the numerical settings and code used for our RHD jet simulations of jets expanding into a homogeneous circumburst medium in \S \ref{simulation_section}. Our approach is made possible by two properties of decollimating and decelerating relativistic jets starting from a BM solution. First, the jet evolution is scale-invariant under rescaling of both explosion energy and circumburst density, which we discuss in \S \ref{scale_invariance_section}. Secondly, for a given initial opening  angle, the 2D fluid profile of the blast wave evolves smoothly from a relativistic and purely radial outflow to a non-relativistic and spherical outflow. This implies that both the radial and lateral structure of the flow can be captured with sufficient accuracy by a low resolution grid with specialized coordinates that can be determined \emph{a posteriori} once the radial and angular extent of the jet at each moment in lab frame time are known from the high-resolution simulation. The dynamical evolution and condensed low-resolution  description of jets with various opening angles are discussed in \S \ref{jet_dynamics_section}. In \S \ref{box_fitting_section} we describe how the simulation results have been implemented in a broadband fitting code, which we apply to a case study, i.e. GRB 990510, in \S \ref{GRB990510_section}. We discuss our results in \S \ref{discussion_section}.

The source code of the broadband fit code will be made freely available on request or for download from \url{http://cosmo.nyu.edu/afterglowlibrary}. It can be run both on a single core and in parallel, and allows the user to either quickly generate light curves and spectra for arbitrary explosion parameters, or, when provided with a data set of observed fluxes in mJy, to perform a full broadband fit.

\section{Numerical jet simulations}
\label{simulation_section}

For this study a total of 19 jet simulations in 2D have been performed using the Relativistic Adaptive Mesh (\textsc{RAM}) parallel RHD code \citep{Zhang2006}. The code employs the fifth-order weighted essentially non-oscillatory (WENO) scheme \citep{Jiang1996} and uses the PARAMESH AMR tools \citep{MacNeice2000} from FLASH 2.3 \citep{Fryxell2000}. For all jet simulations the BM solution for an adiabatic impulsive explosion is used in spherical coordinates to set the initial conditions. Instead of the full spherical solution, a conic section is used that is truncated at a different fixed opening angle for each simulation. The opening angles are listed in table \ref{sim_table}, along with the jet energy $E_j$ for each simulation.

\begin{table}[h]
\centering
\caption{Opening angles $\theta_0$ in radians and jet energies $E_j$ in erg for each simulation.}
\begin{tabular}{l|ll}
 & $\theta_0$ (rad) & $E_j$ (erg) \\
\hline
 1 & 0.045 & $6.328 \cdot 10^{48}$ \\
 2 & 0.05  & $7.812 \cdot 10^{48}$ \\
 3 & 0.075 & $1.758 \cdot 10^{49}$ \\
 4 & 0.1   & $3.125 \cdot 10^{49}$ \\
 5 & 0.125 & $4.883 \cdot 10^{49}$ \\
 6 & 0.15  & $7.031 \cdot 10^{49}$ \\
 7 & 0.175 & $9.570 \cdot 10^{49}$ \\
 8 & 0.2   & $1.250 \cdot 10^{50}$ \\
 9 & 0.225 & $1.582 \cdot 10^{50}$ \\
10 & 0.25  & $1.953 \cdot 10^{50}$ \\
11 & 0.275 & $2.363 \cdot 10^{50}$ \\
12 & 0.3   & $2.813 \cdot 10^{50}$ \\
13 & 0.325 & $3.301 \cdot 10^{50}$ \\
14 & 0.35  & $3.828 \cdot 10^{50}$ \\
15 & 0.375 & $4.395 \cdot 10^{50}$ \\
16 & 0.4   & $5.000 \cdot 10^{50}$ \\
17 & 0.425 & $5.645 \cdot 10^{50}$ \\
18 & 0.45  & $6.328 \cdot 10^{50}$ \\
19 & 0.5   & $7.812 \cdot 10^{50}$ \\
\end{tabular}
\label{sim_table} 
\end{table} 

The jet energy $E_j$ (the total for both jets) relates to the isotropic equivalent energy $E_{iso}$ according to
\begin{equation}
 E_j = E_{iso} (1 - \cos \theta_0 ) \approx E_{iso} \theta_0^2 / 2.
\end{equation}
All jets expand into a homogeneous medium with number density $n_0 = 1$ cm$^{-3}$ (mass density $\rho_0 = 1 \times m_p$ g cm$^{-3}$, in terms of the proton mass $m_p$), and have an isotropic equivalent explosion energy $E_{iso} = 6.25 \times 10^{51}$ erg; note that due to the scale invariance of the simulations with respect to $n_0$ and $E_{iso}$ these can be scaled afterwards to represent arbitrary values (see Section \ref{scale_invariance_section}). All jets start at time $t_b$ with fluid Lorentz factor $\gamma_b = 25$ directly behind the shock, ensuring that for all simulations $\gamma_b > 1 / \theta_0$. At this point the edges of the jet have not yet come into causal contact and lateral spreading has not yet set in, which allows us to use the spherically symmetric BM solution as the starting point.

The initial outer radii of the jets are given by $R_b = 1.3102 \times 10^{17}$ cm, determined from $t_b$ and $\gamma_b$ through $R_b = c t_b (1 - 1 / 16 \gamma_b^2)$ (eq. 26 from BM, with $c$ the speed of light). The initial time $t_b$ is determined from $\gamma_b$, $E_{iso}$ and $n_0$ using
\begin{equation}
 17 E_{iso} = 16 \pi m_p n_0 \gamma_b^2 c^5 t_b^3,
\label{energy_equation}
\end{equation}
which expresses conservation of total energy (eq. 43 from BM). 

For all simulations the stopping time is determined according to
\begin{equation}
t_f = 10 \times t_\textrm{NR} = 9700 \left( \frac{E_{iso}}{10^{53} n_0} \right)^{1/3} \textrm{days}
\end{equation}
The time $t_\textrm{NR}$ marks the time when the jet is analytically expected to transition from relativistic to non-relativistic flow, and is determined from comparing the initial explosion energy to the total rest mass energy of the swept-up matter \citep{Piran2005}. Numerical simulations have shown that in practice this transition takes more time to complete, which is why we have chosen to continue our simulations until two times the transition duration $5 \times t_\textrm{NR}$ found numerically by \cite{Zhang2009}. For $E_{iso} = 6.25 \times 10^{51}$ erg and $\gamma_b = 25$, it follows for all 19 simulations that $t_b = 4.37 \cdot 10^6$ s $= 50.6$ days and $t_f = 3.33 \cdot 10^8$ s $= 3849$ days; again, these values change under rescaling of $E_{iso}$ or $n_0$.

All simulations use an equation of state with the adiabatic index as a function of comoving density and pressure changing smoothly from 4/3 for a relativistic fluid to 5/3 for a non-relativistic fluid \citep{Zhang2009, Mignone2005}. 

\subsection{Resolution and Refinement}

The initial width of the shell is $\Delta R_b \sim R_b / 2 \gamma_b^2 \approx 1.05 \times 10^{14}$ cm. 
In order to correctly resolve this initial width, we have set the initial peak refinement level to 15, which for a grid running from $0.01 \times R_b$ to $c \times t_f$ and 384 radial cells at the base level, implies a smallest radial cell size of $\delta r = 1.58 \times 10^{12}$ cm, since each increase in refinement level double the effective resolution. There are 32 base level cells in the angular direction, so that $\delta \theta = 9.6 \times 10^{-5}$ rad. Over time, the peak refinement level is gradually decreased until a peak level of 9 (the same approach has been applied in \citealt{Zhang2009}). 

In addition to the global change in peak refinement level, a number of additional manual derefinement strategies have been employed in order to prevent the simulation from devoting too much of its calculation time and memory to resolving the boundary and non-relativistic sideways shock between the interstellar medium (ISM) and empty region far behind the shock front, as well as the Kelvin-Helmholtz instabilities that arise in the inner low density regions of the shock due to velocity shear at the edge. These regions have no relevance for the jet dynamics or the observed radiation, which is produced close to the front of the shock. The shock front and its sideways expansion are fully resolved. The additional manual derefinement settings are an increasing inner radius behind which derefinement is enforced. This region expands as $r = 1.2 \times 10^{17} (t / t_b)$ until $\gamma = 2$ (according to BM) and then stops. Regions where the fluid number density is below $0.75\,n_0$ have their peak refinement level reduced by 6. The peak level is also reduced by 6 where the local $\gamma < 1.5$, but is never allowed to drop below 9 for derefinement based on this criterion. Finally, the peak level is reduced by 4 where the local $\gamma < 3$, but is never allowed to drop below 11 for derefinement based on this criterion.

\section{Scale invariance of the jet}
\label{scale_invariance_section}

What is straightforward, but perhaps obscured by the richness of features in the resulting light curves, and what has so far not been utilized in afterglow modeling, is that the full 2D evolution of the jet is invariant under scaling of the initial explosion energy and circumburst medium density: independent of self-similarity, a more energetic jet (or a jet in a less dense environment) goes through exactly the same evolutionary stages as a jet with lower energy (or higher circumburst density), albeit that each stage occurs at a later time and larger radius.

The scalings can be understood as follows. The initial set-up of the problem is determined completely by a limited number of parameters: $E_{iso}$, $\rho_0$ ($\equiv n_0 \times m_p$, with $m_p$ the proton mass), $\theta_0$ and $c$. The initial Lorentz factor $\gamma_b$ is not included in the list because its precise value is arbitrary as long as $\gamma_b > 1 / \theta_0$. Only a limited number of independent dimensionless combinations of these parameters is possible, such as:
\begin{equation}
 A = \frac{r}{ct}, \quad B = \frac{E_{iso} t^2}{\rho_0 r^5}, \quad \theta, \quad \theta_0.
\end{equation}
Any dimensionless quantity that describes the local fluid conditions or global evolution of the jet, such as Lorentz factor $\gamma$, density $\rho / \rho_0$ or $\theta_{95}$ (the angle with respect to the jet axis within which $95 \%$ of the jet energy is contained) can be expressed as a function of these parameters, i.e. $\gamma(r,t, \theta, E_{iso}, \rho_0, \theta_0) \to \gamma( A, B, \theta, \theta_0)$. Note that at both very early times (no lateral flow) and at very late times (spherical flow) there is no dependency on $\theta$ or $\theta_0$, which together with the fact that $A$ becomes a constant value both in the ultra-relativistic BM and non-relativistic ST limits (i.e. $A \to 1$ and $A \to 0$, respectively), accounts for the self-similarity of these limiting cases, with $\gamma(r,t, \theta) \to \gamma(B)$, etc. The parameter $B$ is  identical to the ST self-similarity variable $\xi^{-5}$, and is central to the self-similarity of the BM solution via equation \ref{energy_equation} above (where $r \sim c t$), which fixes $\gamma^2$ and therefore the BM self-similarity variable $\chi(B, \gamma^2(B))$.

Even if we do not limit ourselves to either of the self-similar extremes and leave $A$ and $\theta$ in place, we have a set of coordinates $A, B, \theta$ that is invariant under the following rescalings:
\begin{eqnarray}
 E'_{iso} & = & \kappa E_{iso}, \nonumber \\
 \rho_0' & = & \lambda \rho_0, \nonumber \\
 r' & = & (\kappa / \lambda)^{1/3} r, \nonumber \\
 t' & = & (\kappa / \lambda)^{1/3} t.
\end{eqnarray}
The full scale-invariant hydrodynamics equations in terms of $A$, $B$, $\theta$ are provided for completeness in appendix \ref{scale_invariance_appendix}. The direct implication of this is that we can determine the value of any dimensionless quantity for an explosion with $E'_{iso}$ and $\rho'_0$ simply by probing a simulation with parameters $E_{iso}, \rho$ at $t, r$, rather than at $t',r'$. Quantities that are not dimensionless, such as density $\rho$ and internal energy density $e$ follow from $\rho / \rho_0 = \rho' / \rho_0'$, $e / \rho_0 c^2 = e' / \rho_0' c^2$, etc.

\begin{figure}
 \centering
 \includegraphics[width=1.0\columnwidth]{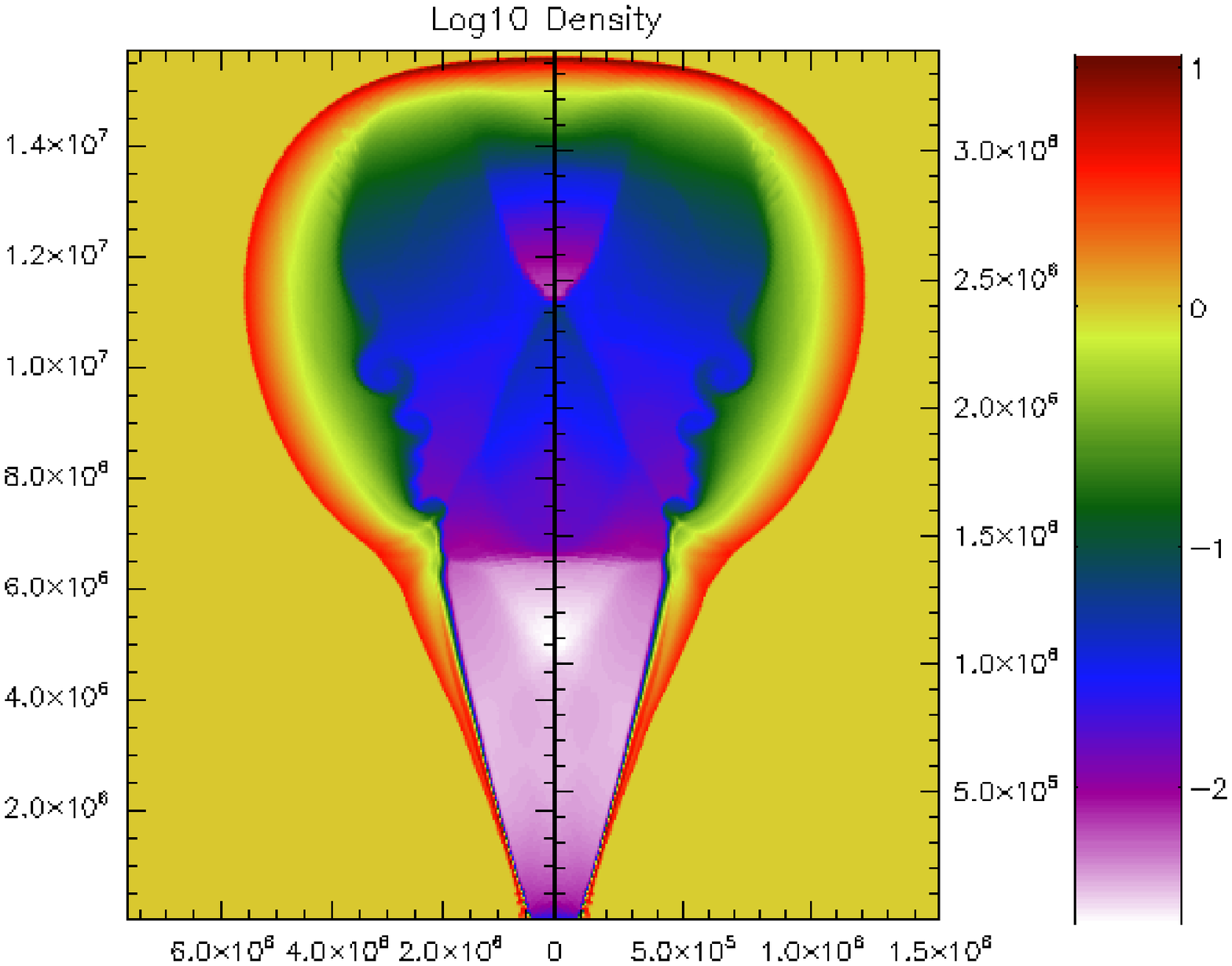}
 \includegraphics[width=1.0\columnwidth]{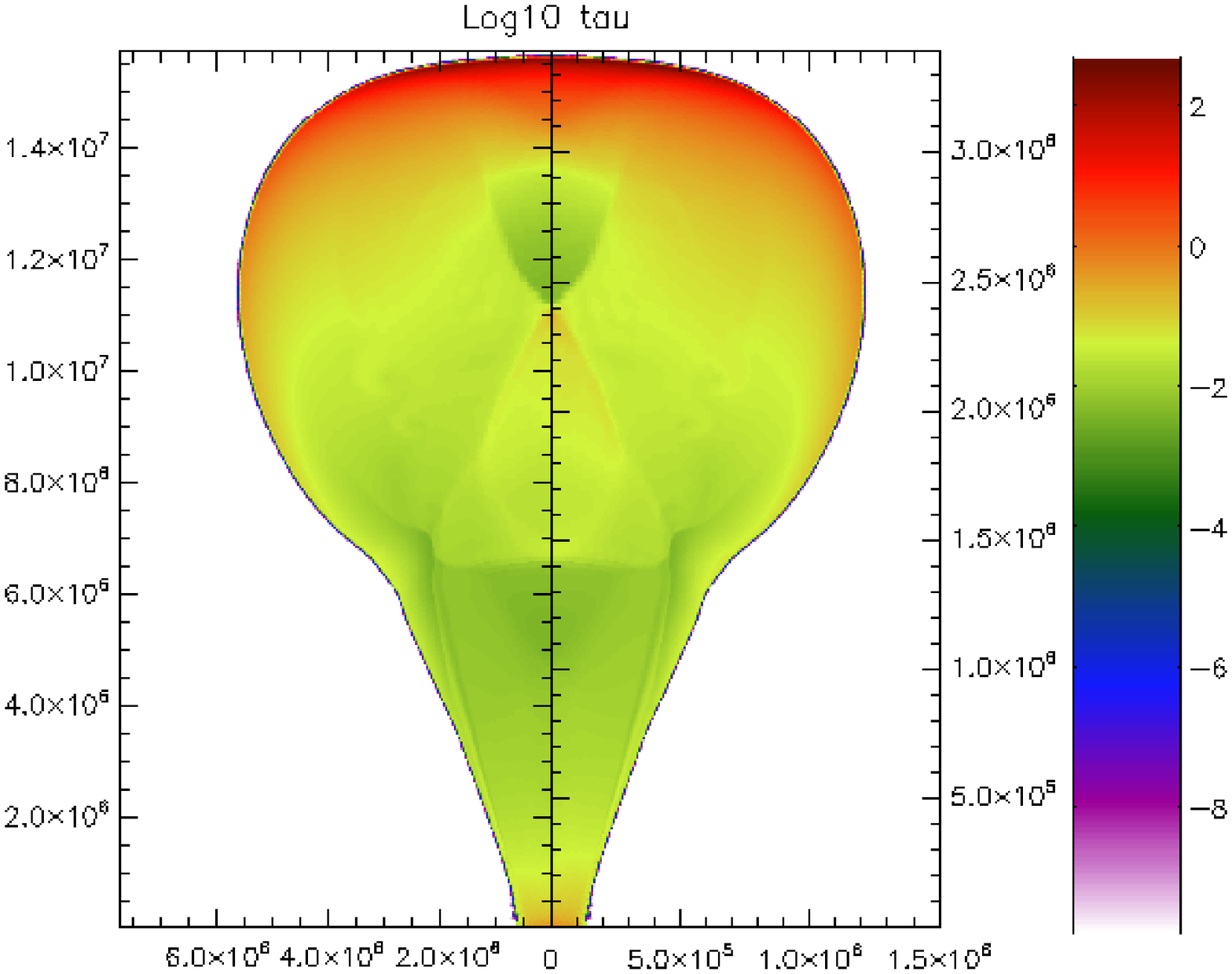}
 \caption{Direct comparison between comoving number density $n$ in cm$^{-3}$ (top) and lab frame energy density $\tau$ in units of $m_p c^2$ (bottom) profiles for jet simulations with $\theta_0 = 0.2$ rad, $n_0 = 1$ cm$^{-3}$ , and $E_{iso} = 5 \times 10^{51}$ erg (left) or $E_{iso} = 5 \times 10^{49}$ erg (right), drawn from \cite{vanEerten2011sgrbs}. The snapshot times differ by a factor $100^{1/3}$ and for both snapshots the fluid Lorentz factor directly behind the shock is $\sim 3.3$.}
 \label{same_density_comparison_figure}
\end{figure}

\begin{figure}
 \centering
 \includegraphics[width=1.0\columnwidth]{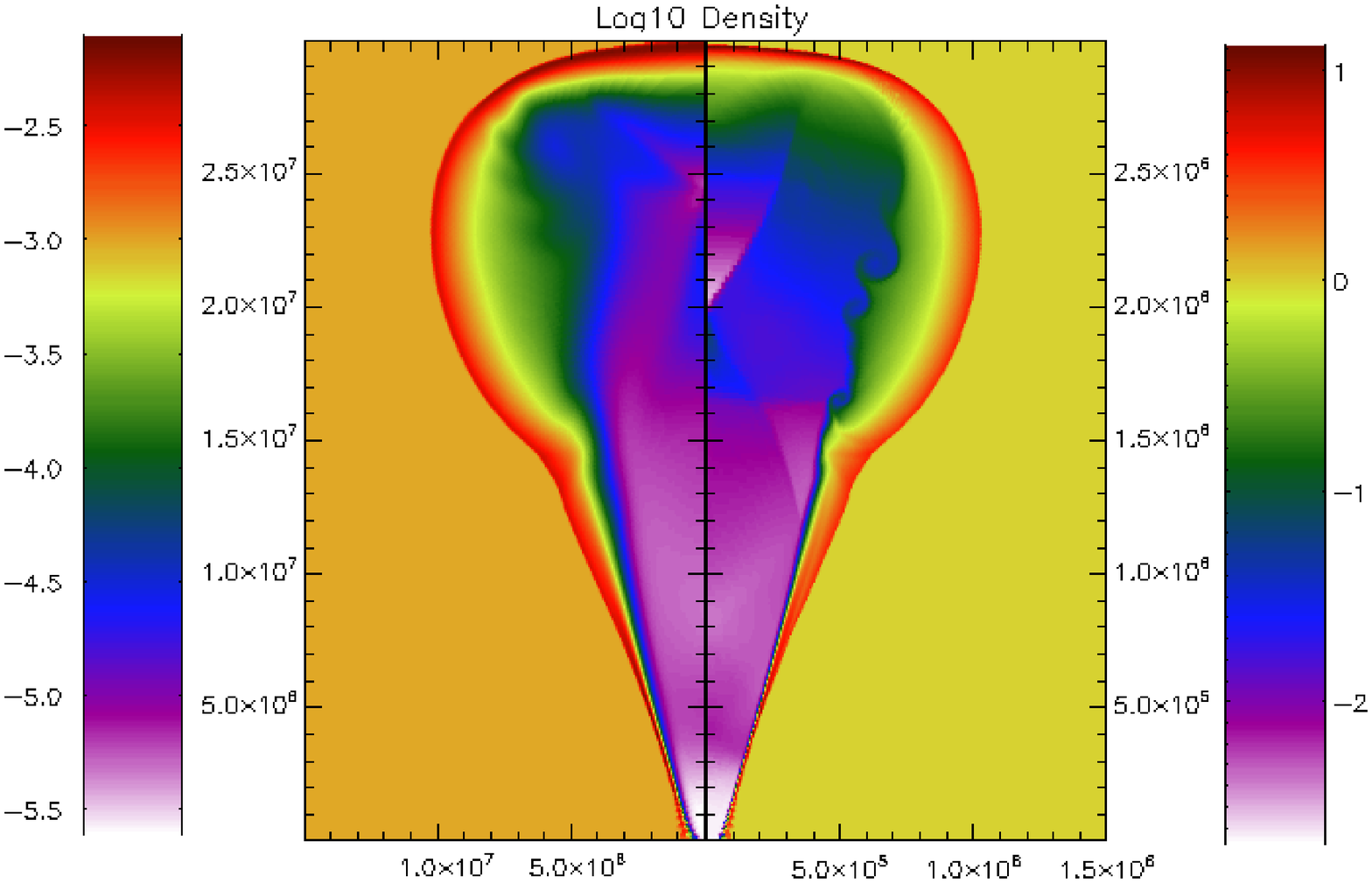}
 \includegraphics[width=1.0\columnwidth]{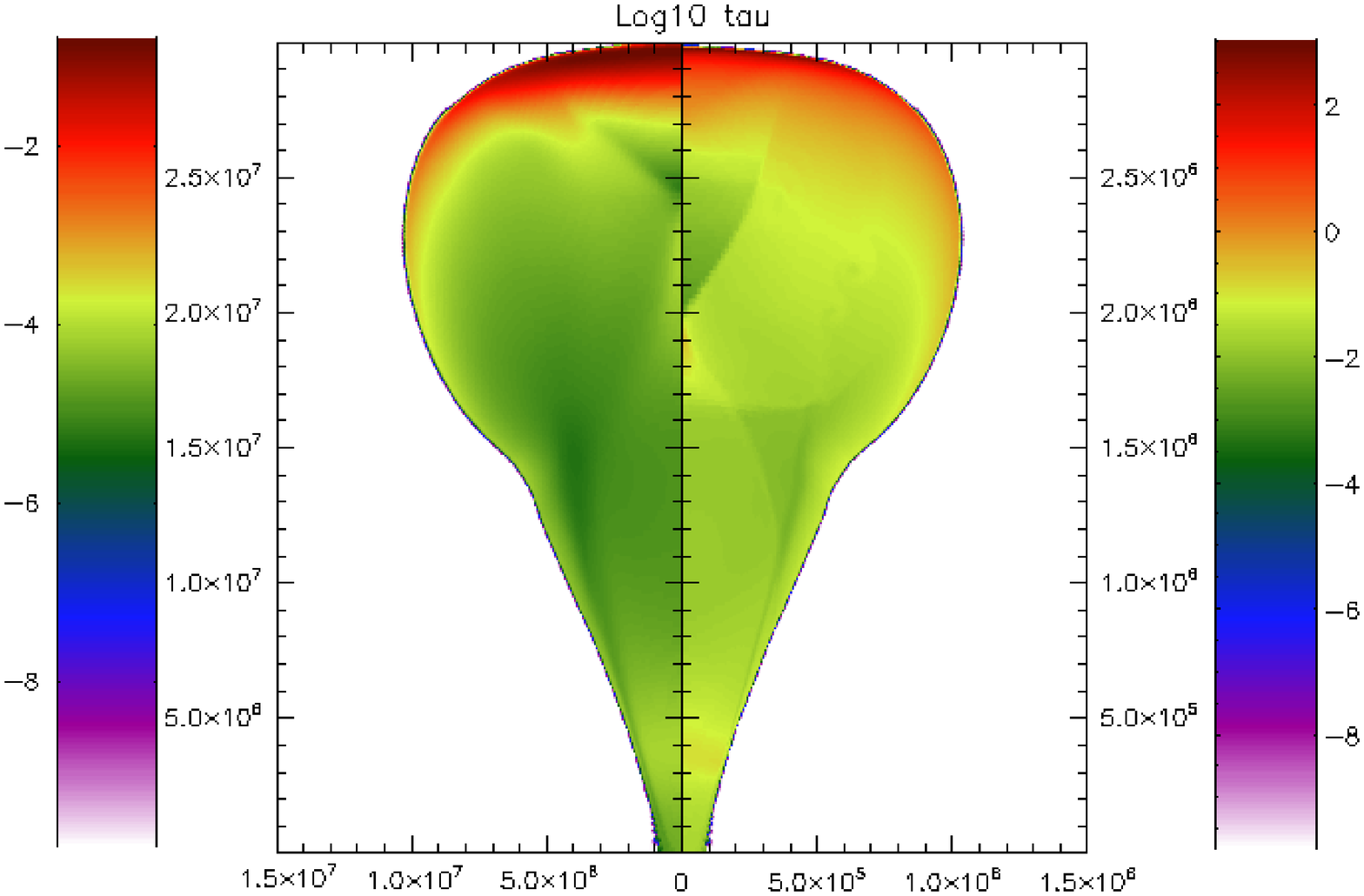}
 \caption{Direct comparison between comoving number density $n$ in cm$^{-3}$ (top) and lab frame energy density $\tau$ in units of $m_p c^2$ (bottom) profiles for jet simulations with $\theta_0 = 0.2$ rad, $E_{iso} = 5 \times 10^{49}$ erg, and $n_0 = 1\times 10^{-3}$ cm$^{-3}$ (left) or $n_0 = 1$ cm$^{-3}$ (right), drawn from \cite{vanEerten2011sgrbs}. Unlike in the case of the two simulations of Figure \ref{same_density_comparison_figure}, there is now a scaling factor $\lambda = 10^3$ between fluid quantities $\rho$ and $\tau$ for the different fluids as well. For both snapshots the fluid Lorentz factor directly behind the shock is $\sim 4$.}
 \label{different_density_comparison_figure}
\end{figure}

Examples of scalings between jets with different $E_{iso}$ and $\rho_0$ are given in Figures \ref{same_density_comparison_figure} and \ref{different_density_comparison_figure}. The comoving fluid density and energy density profiles are drawn from the simulations presented in \cite{vanEerten2011sgrbs}, since the 19 simulations listed in table \ref{sim_table} were set up to be unrelated via scaling. In the case of Figure \ref{same_density_comparison_figure}, two simulations are compared for which $\lambda = 1$, whereas $\lambda = 10^3$ in the case of Figure \ref{different_density_comparison_figure}. 
In Figure \ref{same_density_comparison_figure} we set the outer radius of the left and right panels equal to their respective $c t$, resulting in both images being complete mirror images of each other, which confirms that both simulations are numerically indistinguishable. Although analytically equivalent, the two simulation runs in Fig. \ref{different_density_comparison_figure} are no longer numerically identical, which leads to minor differences in (de-)refinement in the inner regions. However, those inner regions do not contribute to the observed flux, and thus have no observable effect on the light curves.

\section{Lateral spreading and jet deceleration}

\label{jet_dynamics_section}
\begin{figure}
 \centering
 \includegraphics[width=1.0\columnwidth]{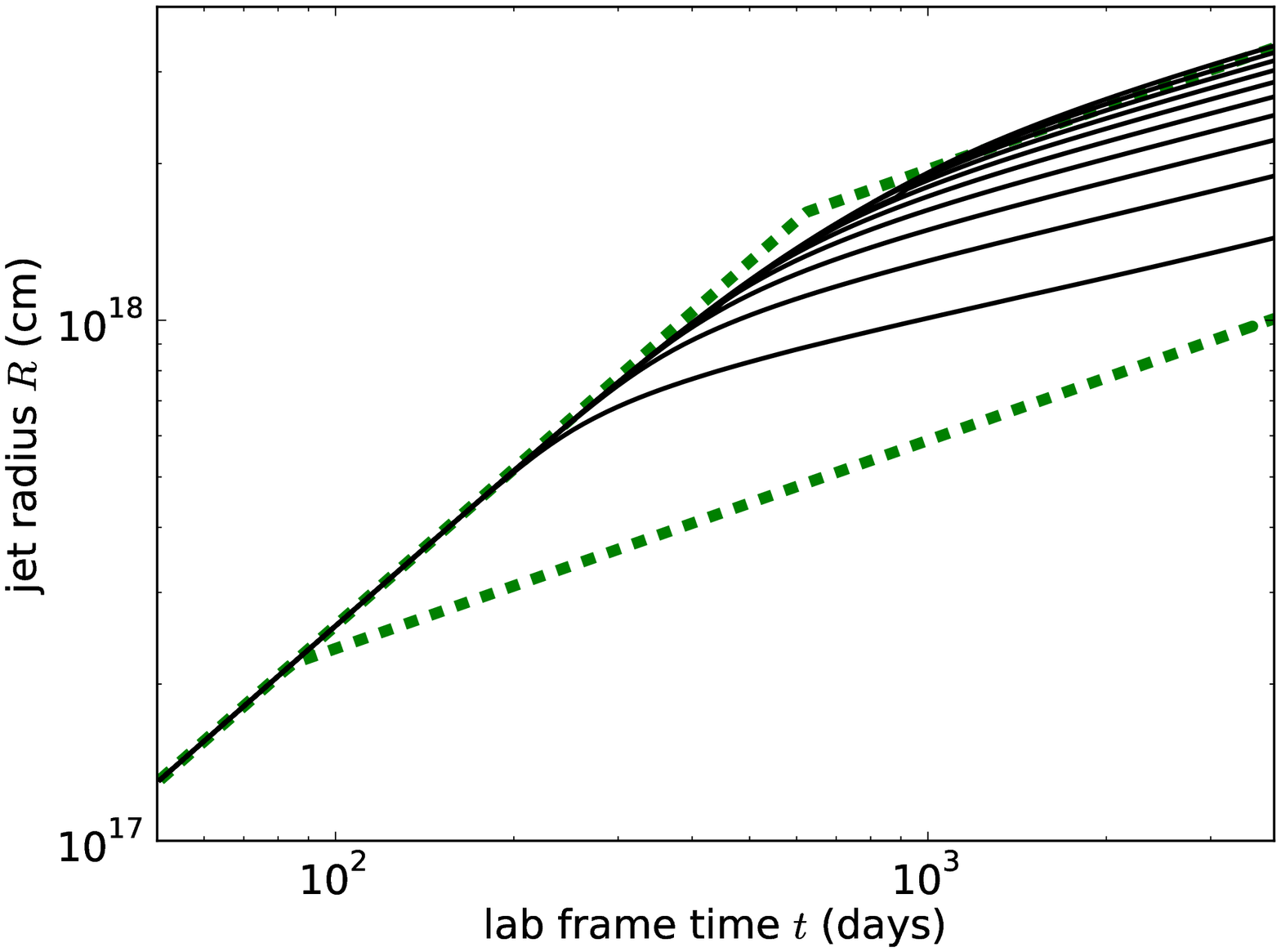}
 \includegraphics[width=1.0\columnwidth]{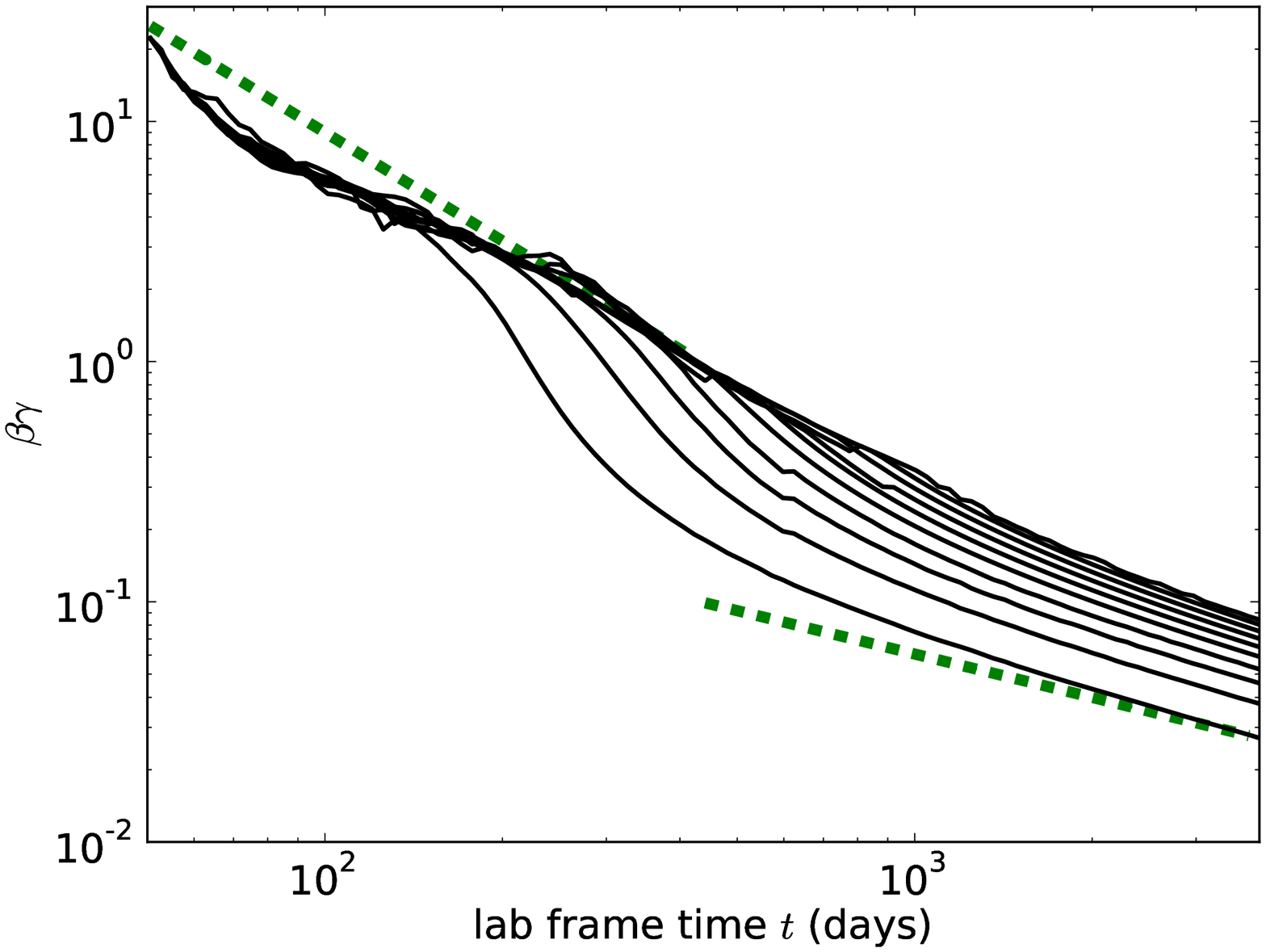}
 \caption{Top: evolution of the blast wave radius $R$ over lab frame time for jet with $\theta_0 =$ 0.5, 0.45, 0.4, 0.35, 0.3, 0.25, 0.2, 0.15, 0.1, 0.05 rad (top to bottom in both plots). Dashed green lines indicate asymptotic BM and ST predictions, both for a spherical blast wave and for $\theta_0 = 0.05$ rad. Bottom: evolution of blast wave velocity $\beta \gamma$ for the same opening angle jets. Early time dashed green line indicates BM prediction of $\beta_r \gamma_r \propto t^{-3/2}$, late time indicates $\beta_r \gamma_r \propto t^{-3/5}$.}
 \label{blast_wave_radius_figure}
\end{figure}

\begin{figure}
 \centering
 \includegraphics[width=1.0\columnwidth]{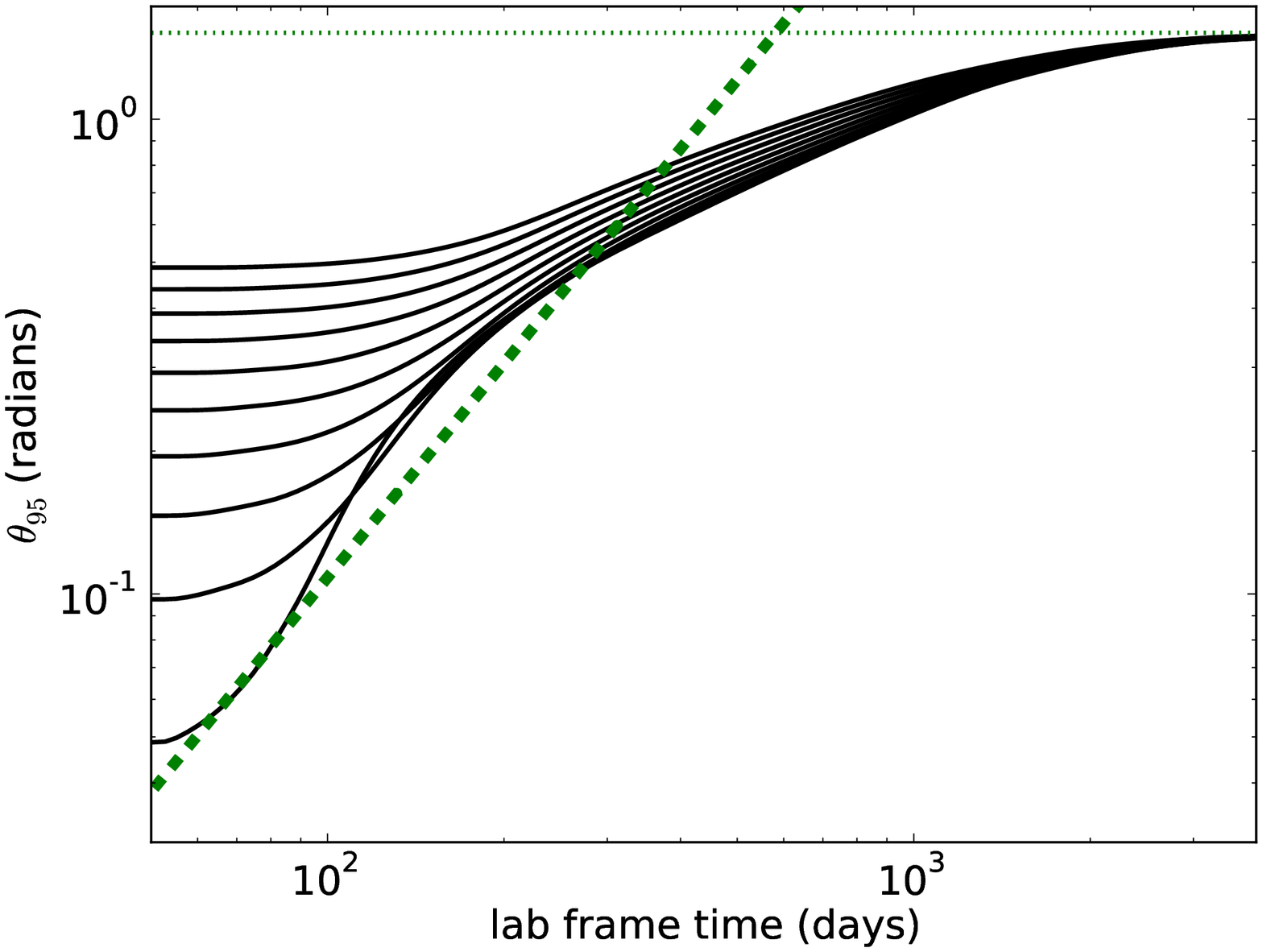}
 \includegraphics[width=1.0\columnwidth]{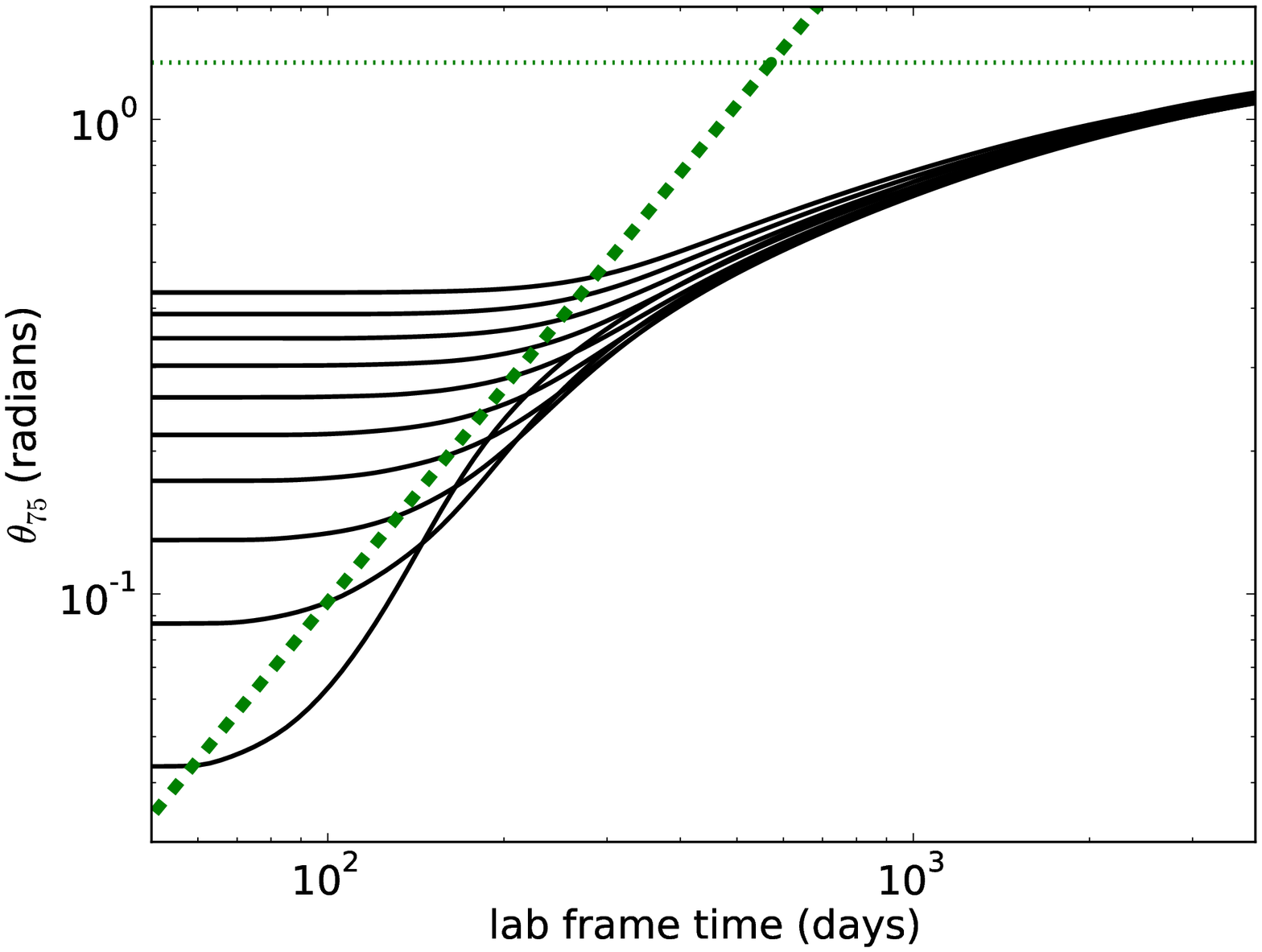}
 \includegraphics[width=1.0\columnwidth]{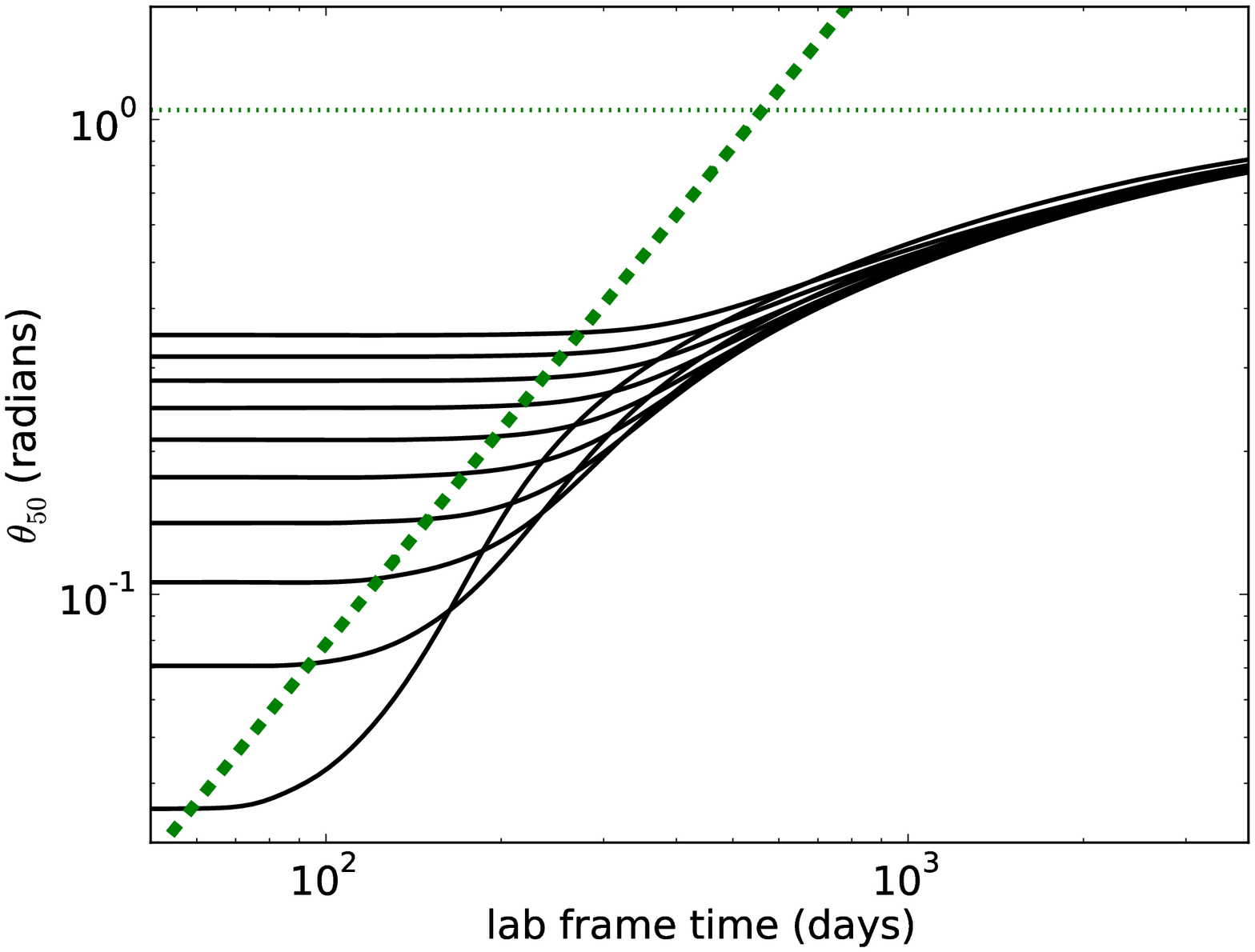}
 \caption{Lateral spreading of jet simulations with opening angles $\theta_0 =$  0.5, 0.45, 0.4, 0.35, 0.3, 0.25, 0.2, 0.15, 0.1, 0.05 rad (top to bottom in each plot). The top plot shows the evolution of $\theta_{95}$, defined as the angle within which 0.95 of the volume-integrated rest frame energy density $\tau$ is contained. Center plot shows the same for $\theta_{75}$ (0.75 of energy) and the bottom plot for $\theta_{50}$ (0.50 of energy). Dashed green lines indicate when $\gamma = 1 / \theta_0$ for the different opening angles (shifted to correct for fractions $< 1$) and should therefore be compared to the initial opening angles of the jets, not to $\theta_{xx}$ angles at any given time. The horizontal dotted green lines indicate the values for the limiting angles for the case of a homogeneous spherical blast wave.}
 \label{theta_max_figure}
\end{figure}

We have plotted a number of features of a subset of the 19 jet simulations in Figures \ref{blast_wave_radius_figure} and \ref{theta_max_figure}. The jets follow the same evolution as those of earlier of studies, such as \cite{Zhang2009, vanEerten2011jetspreading, Wygoda2011}. The top plot of Figure \ref{blast_wave_radius_figure} shows the evolution of the on-axis outer radius of the blast wave for jets with different $\theta_0$ in the lab frame of the explosion. All jet simulations start out relativistically with identical $\gamma_b$, $t_b$ and blast wave radius $R_b$. Jets with a wide opening angle do not have a long spreading phase and undergo a single smooth transition from $R \approx c t$ to $R \approx 1.15 (E_j t^2 / \rho_0)^{1/5}$ (where the numerical factor 1.15 follows from the ST solution with adiabatic index $5/3$). The figure shows that for a very wide jet with $\theta_0 = 0.5$, the transition time is well approximated by the crossing point of the asymptotes for spherical outflow. However, this does not carry over to smaller angle jets, and the meeting point of the asymptotes for $E_j (\theta_0 = 0.05)$ severely underestimates the turnover point. The reason for this is that for narrower jets there is also an intermediate phase, where the jet decelerates due to lateral spreading. Although not as abrupt as originally predicted (\citealt{Rhoads1999}, also discussed in \citealt{vanEerten2011jetspreading, Wygoda2011}) and occurring throughout the transrelativistic phase of the jet, this leads to an extended period of jet deceleration in excess of the asymptotic jet deceleration of the ST phase and adds a second turnover point in the evolution of jet radius and jet velocity. As the bottom plot of Figure \ref{blast_wave_radius_figure} indicates, this intermediate phase does not follow a simple power law. A full parameterization of the intermediate phase lies beyond the scope of this work and will not be required for model fitting based on the simulation results. In addition, due to inhomogeneity along the shock front, deceleration will be different for outflows along different angles. 

The early time behavior visible in the bottom plot of Figure \ref{blast_wave_radius_figure}, where the peak Lorentz factor initially drops below its expected value in the BM regime but then moves back to the BM asymptote, is due to the resolution of the simulations. In the BM solution, the blast wave is extremely thin ($\Delta R_b \sim R_b / 2 \gamma^2_b)$ and we have not been able to achieve full convergence in 2D at early times (the issue is identical to that illustrated in Figure 5 of \citealt{Zhang2009}). By using the integrated values of the fluid quantities across a single fluid cell rather than the values at its central coordinate, we have ensured that all energy of the BM solution is accounted for during the initialization of the simulation. For this reason, the drop is only temporary. We emphasize that this is \emph{not} due to lateral spreading of the jet, for if that were the case the peak Lorentz factor would not have been able to recover.

Lateral spreading and the inhomogeneity of the shock front are illustrated in Figure \ref{theta_max_figure}. The plots show the time evolution of various characteristic angles $\theta_{95}, \theta_{75}, \theta_{50}$ of the outflow, defined as the angles within which a fraction $0.95, 0.75, 0.50$ of the volume-integrated rest frame energy density $\tau$ is contained. The green dashed lines indicate the analytically expected onset of lateral spreading, when $\gamma \sim 1 / \theta_0$. Because the plots show $\theta_{95},  \ldots$ rather than the outer edge angle $\theta_{edge}$, these lines have been shifted according to $\theta(t) \to \sqrt{0.95} \theta_{edge}$ (or $\sqrt{0.75} \theta_{edge}$ or $\sqrt{0.50} \theta_{edge}$),  This means that if the simulations had followed the analytical estimate, the turnovers for all angles and fractions would have occurred along the green lines. Because the jets become strongly inhomogeneous along the shock front, the turnover is delayed for characteristic angles bounding smaller energy fractions: the jet energy stays concentrated near the jet axis even after the edges have begun to expand. Because narrower jets have smaller values of $E_j$, they will decelerate earlier, which leads to the behavior shown in the plots where $\theta_{95}, \ldots$ for small jets cross that of large jets with the same $E_{iso}$. Jets with the same $E_j$ do not show this effect (the characteristic angle curves for wider jets would shift sufficiently far to the left in the plots if their energies were downscaled to match the $E_j$ of the narrowest jet). At late times all curves tend to the same fixed fraction of $\pi / 2$ regardless of initial energy, i.e. the jet becomes spherical and homogeneous along the shock front, but only for high energy fractions is this limit actually reached in our simulations. The early time turnover behavior confirms the validity of our choice of starting $\gamma_b > \theta_0$ (note that for the wide jets $\gamma_b \gg \theta_0$).

\section{``Box''-based broadband afterglow fitting}
\label{box_fitting_section}

Each of the 19 simulations generates 3000 snapshots, varying in data size between $\sim 350$MB at early times and $\sim 40$MB at late times (when lower peak refinement levels are utilized). Therefore it is currently not practically possible to load the complete output for a single simulation into computer memory at once, let alone the combined output of all 19 simulations. However, if we wish to use the simulations as a basis for iterative model fitting, we will need to be able to quickly access the fluid state at any requested point in time and space. We therefore need to summarize the simulation results in a way that adequately captures all aspects of the outflow but occupies only a relatively small amount of computer memory. In this study we have aimed for $< 1$GB in total, but the desired size in memory will be hardware-dependent.

The next three subsections describe the further technical aspects of producing light curves from the summarized `box', rather than the `grid' which originally contains the simulation data. In subsection \ref{fit_methods_section} we describe the details of the fit method.

\subsection{``Box'' summary of simulations}

The phase space for each fluid variable is set by the variables $E_{iso}$, $n_0$, $\theta_0$, $r$, $\theta$ and $t$ and the local fluid state is fully specified by the fluid variables $\tau$, $\rho$, $v_r$ (radial fluid velocity), $v_\theta$ (angular fluid velocity), from which the other fluid quantities such as $p$ (pressure), $e$ or $\gamma$ can be derived. By storing $e$ explicitly along with $\rho$, $v_r$, $v_\theta$, while still including $\tau$, even though the latter is not needed for the radiative transfer calculation, we ensure that we can explore the full fluid state at each point in time and space without having to use the equation of state. This implies that we will have to store five fluid variables along with the coordinates and sizes of the fluid cells they refer to (i.e. 9 quantities in total). The scale invariance discussed in section \ref{scale_invariance_section} implies that we only need to store a single value for $E_{iso}$ and $n_0$. As mentioned already, we have used $19$ possible values for $\theta_0$. We calculate light curves for intermediate values of $\theta_0$ by interpolating the appropriate results from the 19 simulations, as we demonstrate in subsection \ref{light_curves_subsection}. For the $r$, $\theta$ coordinates we use 100 entries each. For $t$ we also use 100 entries rather than 3000, and as with the small subset of $\theta_0$, we demonstrate in subsection \ref{light_curves_subsection} that this number is sufficient and that the full light curve can be calculated using interpolation. All together, we now have the following number of entries in our summary of the fluid evolution:
\begin{math}
E_{iso} \times n_0 \times \theta_0 \times r \times \theta \times t \times variables = 1 \times 1 \times 19 \times 100 \times 100 \times 100 \times 9 = 171,000,000.
\end{math}
We will refer to this summary as the ``box'', in order to distinguish it from grid based fluid profiles from the simulations.

The reason that we only need as few as 100 cells in the $r$ and $\theta$ directions has already been demonstrated partially by Figures \ref{blast_wave_radius_figure} and \ref{theta_max_figure} in the preceding section. These Figures illustrate that, although high-resolution simulations were required in order to accurately determine the blast wave radius and lateral extent at each point in time, the evolution from BM to ST profile itself is smooth. This means that, once the large-scale properties of the outflow are known, it is possible to use this knowledge to select at each point in time new coordinates such that the key features of the outflow are resolved while ignoring parts of the outflow.

\begin{figure}[h]
 \centering
 \includegraphics[width=1.0\columnwidth]{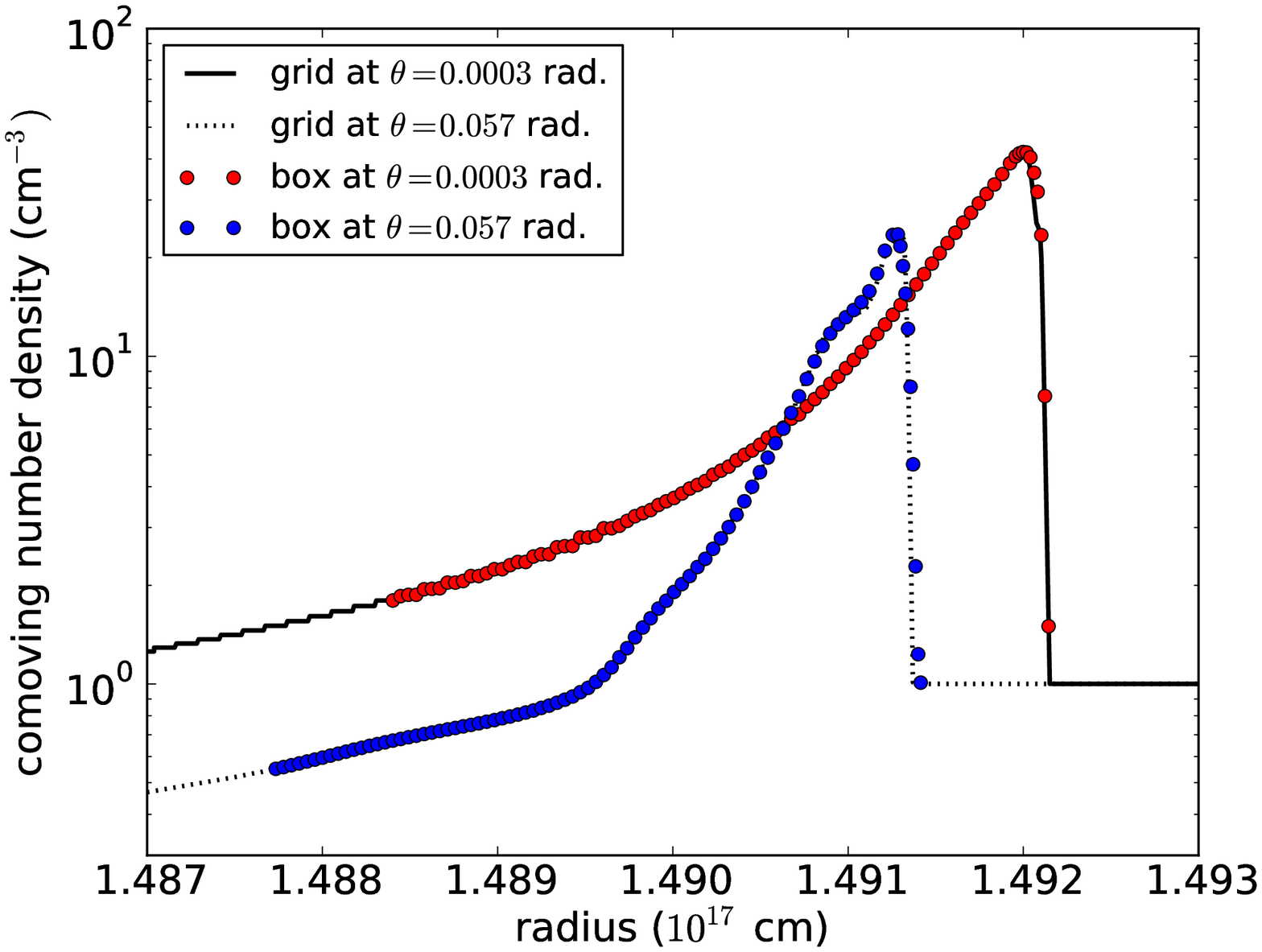}
 \includegraphics[width=1.0\columnwidth]{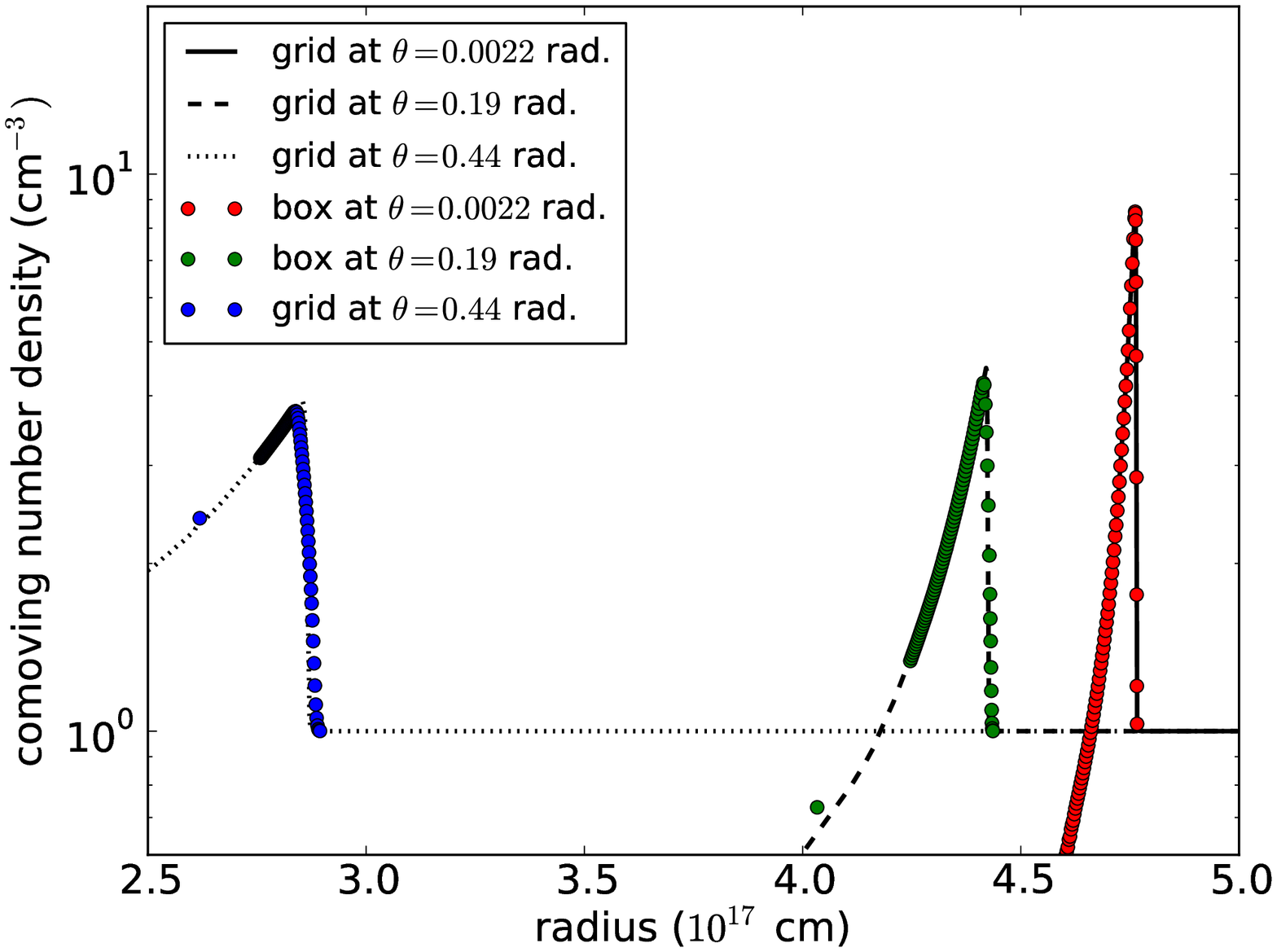}
 \caption{Comoving number densities $n$ for the $\theta_0 = 0.05$ rad simulation at lab frame times $t = 57.6$ days (top) and $t = 186$ days (bottom). At these times the BM solution predicts respectively $\gamma = 20.6$ and $\gamma = 3.6$ behind the shock front (see also Figure \ref{blast_wave_radius_figure}). The fluid profiles from the grid at the indicated angles are drawn with lines, while the box data points are presented by circles. The angles represent values close to the jet center, edge and, in the bottom plot, halfway between center and edge. The grid values are taken exactly at the listed angles, while the box values are averages centered on these angles with $\delta \theta = 5.75 \times 10^{-4}$ rad and $\delta \theta = 4.4 \times 10^{-3}$ rad at early and late time respectively. This accounts for small differences between box and grid values. The large $\theta$ profile at the left in the bottom plot differs from the others in that many box cells are used to resolve the steep drop in front of the blast wave as well, which illustrates that the peak values of $n$ and $\tau$ are not numerically identical in fluid simulations (in this case, the peak value of $\tau$ occurs at larger radius than the peak value of $n$).}
 \label{rho_slices_figure}
\end{figure}

The box $\theta$ at each point in time will not cover the entire grid but only runs from 0 to $\theta_{MAX} \equiv \theta_{99} / 0.99$ (i.e. slightly over the the outer angle within which 99 \% of the volume integrated $\tau$ is contained). Anything outside these angles is either ISM or contributes a negligible amount of radiation. The lateral cells within this region are evenly spread. Alternatively we could have set the cell coordinates using $\theta_{00}$, $\theta_{01}$, etc., but there is no significant difference in the resulting light curves. The radial domain runs from $0$ to the outer limit $R_{MAX}$ of the blast wave at each box value of $\theta$. Because the unshocked ISM is at rest, the outer boundary of the shock wave is readily determined using the criterion that $\gamma > 1.000001$. Even for extremely high resolution, this point will not exactly coincide with the peak of the blast wave. We therefore devote 10 cells to the region between $R$ determined by the $\tau$ peak of the blast wave and the first shocked ISM cell. Of the remaining 90 radial box cells, 80 are devoted to resolving the blast wave. Since we know from the BM solution that the blast wave width is $\Delta R \approx R / 12 \gamma^2$ initially and from the ST solution that $\Delta R \propto R$ eventually, we analytically determine the width of the blast wave by $\Delta R \equiv R / 12 \gamma^2$, with $\gamma^2 \equiv \gamma_b (t / t_b)^{-3/2} + 1$ (where the '1' has been added relative to BM eq. 24 to obtain the correct asymptotic behavior). Note that this is only approximately correct on-axis due to 2D spreading and less accurate for the radial profile along the outer angles. This does not matter, however, as long as the approximation is sufficient to resolve the sharp feature of the blast wave. The final ten radial cells are spaced between the origin and the back of the shock. All radial cells are equally spaced within their respective region. Figure \ref{rho_slices_figure} shows radial fluid profiles for the $\theta_0 = 0.05$ simulation for different times and angles. Profiles for other values of $\theta_0$ are similar.

\subsection{``Box'' interpolation}

The previous subsection refers to interpolation in $t$ as well as in $\theta_0$. Even though 100 time snapshots are adequate to capture the dynamics of the jet, evaluating the linear radiative transfer integral at just these 100 emission times (that is, in addition to any evaluations of the BM solution at times before the initial simulation time) has been found to lead to noise in the light curve both at early times and during the rise of the counterjet at late times (corresponding to early emission times for the counterjet). The solution to this is to interpolate the fluid profiles between different emission times, something that is neither difficult nor time consuming when done based on the box snapshots. We have found for light curves with $\theta_0$ values between those in Table \ref{sim_table} that interpolation at the fluid level is more effective than interpolation at the level of the light curve: calculating two light curves for adjacent tabulated $\theta_0$ values and interpolating between them will systematically shift the jet break time and post-break asymptote relative to their actual values for the intermediate $\theta_0$. Therefore, both the time and $\theta_0$ interpolation occur at the same stage.

In practice $\sim 3000$ interpolated times have been found to be more than sufficient to remove the numerical noise in the light curve. Implementing the $\theta_0$ and $t$ interpolations to obtain the local fluid state for requested fluid coordinates $r, \theta, t$ works as follows:
\begin{itemize}
 \item There are four tabulated entries needed for the interpolation, determined by the closest surrounding values around the requested $\theta_0$ and $t$. For each entry, scale $\theta_{MAX}$ by first interpolating in $t$, then in $\theta_0$.
 \item For all four entries, determine the appropriate $R_{MAX} (\theta)$, after the $\theta$ coordinates' outer boundaries have been scaled to their new $\theta_{MAX}$. Interpolate this $R_{MAX}$ in $t$ for both $\theta_0$ options separately.
 \item Obtain the box fluid conditions at $r / (\kappa / \lambda)^{1/3}$, $\theta$ for all four entries, applying both the $\theta_{MAX}$ scaling and $R_{MAX}$ scalings.
\item Multiply all non-dimensionless quantities by $\lambda$.
\item Interpolate the results first in $t$, then in $\theta_0$ to obtain the final value for the fluid quantity.
\end{itemize}
We use linear interpolations, which produces converged results (see \S \ref{light_curves_subsection}).

\subsection{Light curves}
\label{light_curves_subsection}

We calculate light curves and spectra from the box using the same method that we have used previously for grid-based light curve calculations \citep{vanEerten2010c, vanEerten2011sgrbs}. The dominant radiation mechanism is assumed to be synchrotron radiation and the broadband emission from each fluid cell is a given by a series of connected power laws similar to those in \cite{Sari1998}. The linear radiative transfer equations are solved simultaneously for a large number of rays. For each point in lab frame time $t$, the plane perpendicular to the direction of the observer and at a fixed distance $R_{EDS}$ from the origin of the box (or grid), defined by $R_{EDS} / c = t - t_{obs}$, defines the area from which emission will arrive at exactly the same observer time $t_{obs}$. This plane is labeled the equidistant surface (EDS, see also \citealt{vanEerten2010}). In earlier work we have employed a procedure analogous to AMR for dynamically changing the number of rays through the EDSs that are followed simultaneously. For the blast waves of this study all EDS refinement would have occurred near the center of the EDS (defined as the intersection of the line from the grid origin to the observer and the EDS), in order to resolve the early time BM profile and the refinement level would have gradually decreased outwards. Since this is essentially equivalent to base 2 logarithmic spacing, for the current study we use fixed logarithmic spacing between rays in the radial direction instead. The number of rays is evenly spaced in the angular direction. This approach is somewhat faster than the dynamical refining and requires less memory for bookkeeping. In the final step the rays are integrated over to yield the observed flux. Flux, observed frequency and observer time are corrected for redshift $z$ during the calculation.

In Appendix \ref{coefficients_section} we give the exact expressions for the emission and absorption coefficients that are calculated while solving the radiative transfer equations through the evolving fluid. Their local values depend on a number of parameters that capture the microphysics behind the synchrotron radiative process as well as on the local fluid conditions. There are four such parameters: the power law slope $p$ of the shock-accelerated electrons, the fraction $\epsilon_B$ of magnetic energy relative to thermal energy, the fraction $\epsilon_e$ of downstream thermal energy density in the accelerated electrons, and the fraction $\xi_N$ of the downstream particle number density that participates in the shock-acceleration process. By performing broadband fits on afterglow data using the box-based fit method from this paper, these parameters can be determined from the fits for the first time using the full blast wave evolution.

All simulations start at $\gamma_b = 25$. Before this time the outflow can be described by a conic section of the spherically symmetric BM solution. Because the observed flux at a given observer time contains emission from a wide range of emission times, initially including times for which $\gamma > \gamma_b$, the BM solution is used directly to determine the initial emission and absorption coefficients. Probing the BM solution between $\gamma = 200$ and $\gamma_b$ using 1600 logarithmically spaced emission times has been found to be more than sufficient to capture the early time emission.

\begin{figure}
 \centering
 \includegraphics[width=1.0\columnwidth]{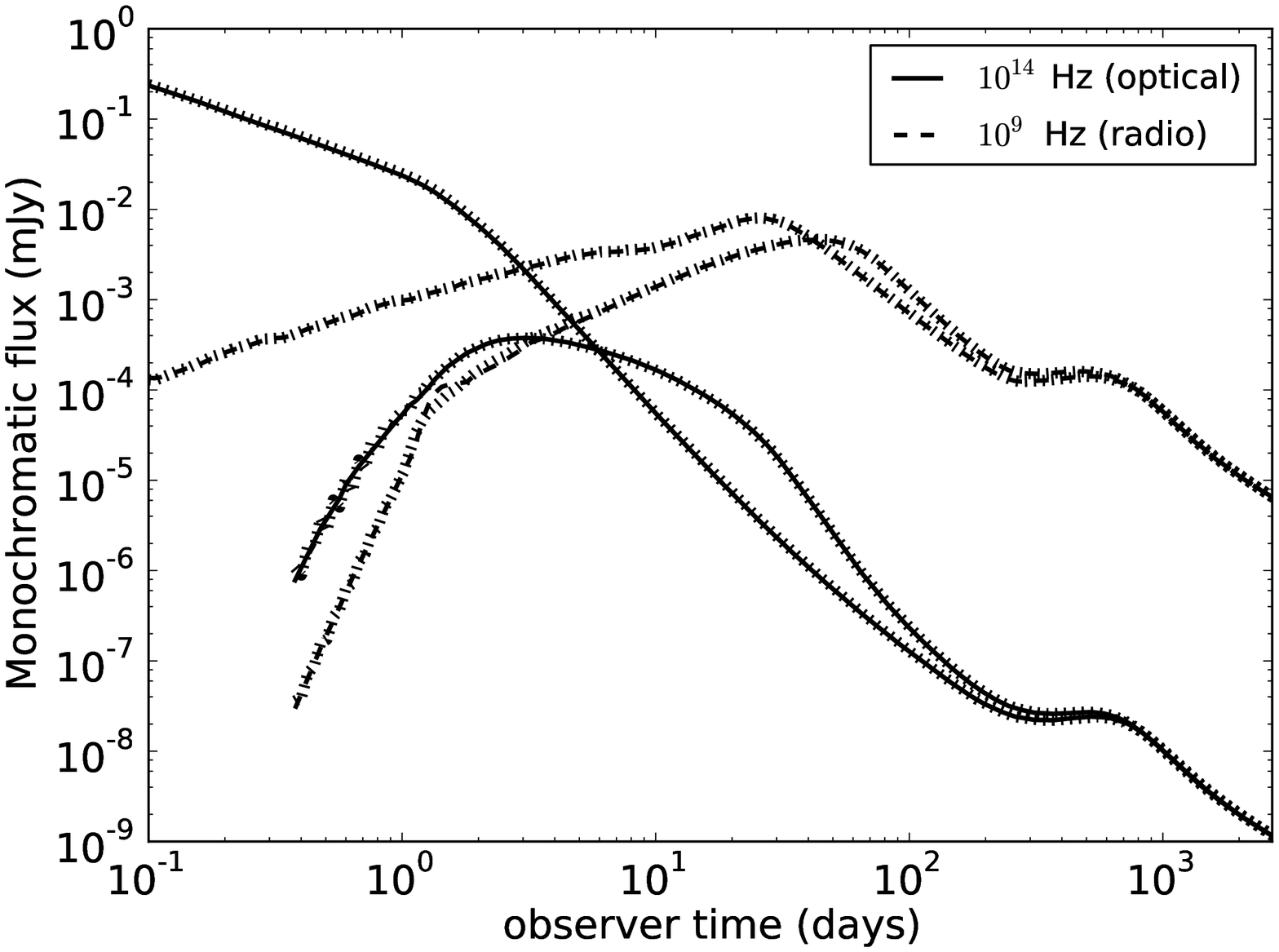}
 \includegraphics[width=1.0\columnwidth]{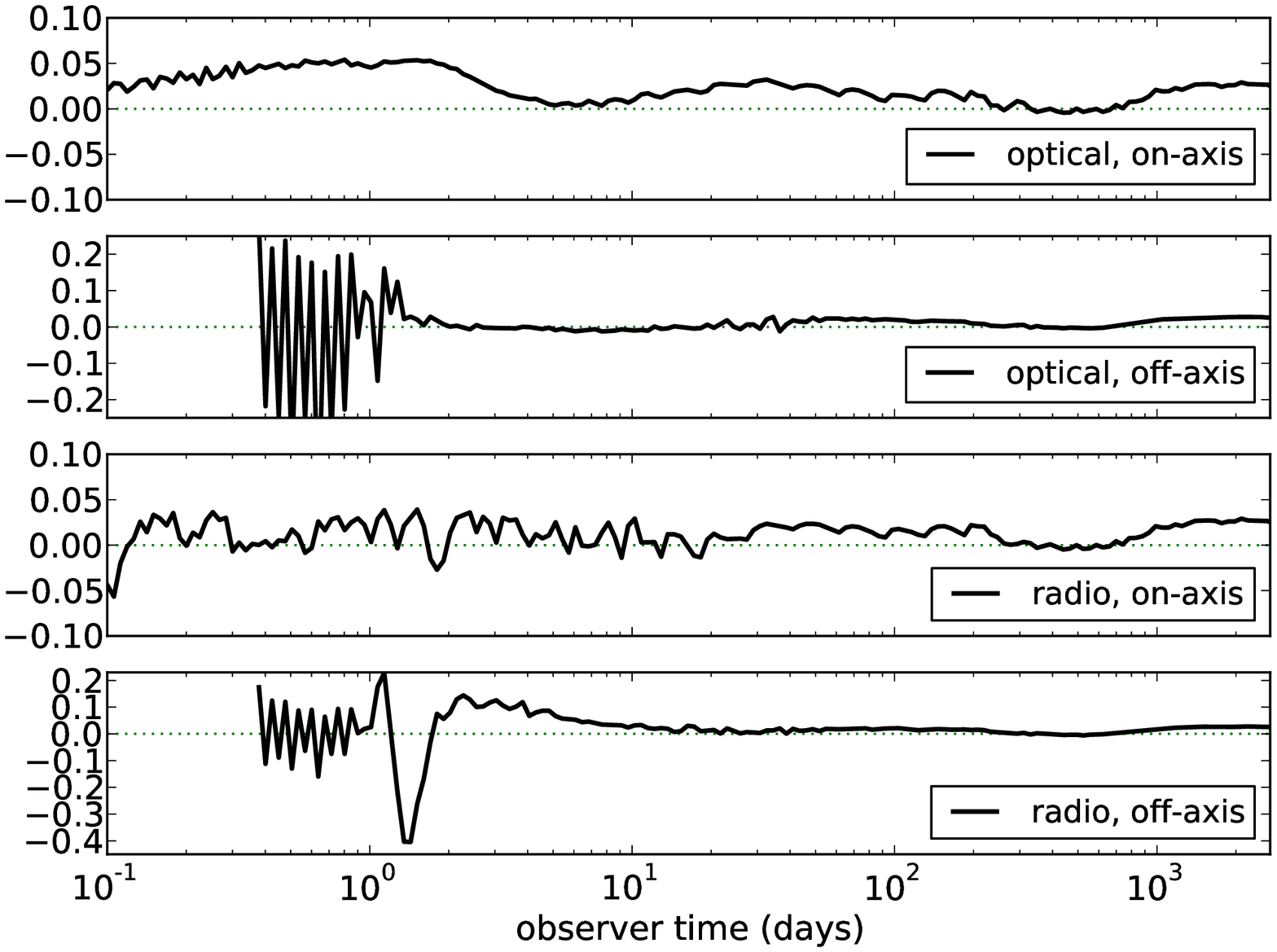}
 \caption{Direct comparison between box light curves and simulation light curves for simulation 8 ($\theta_0 = 0.2$ rad) of this paper. In the top plot, box based curves are shown both for optical (solid lines) on-axis and off-axis ($\theta_{obs} = 0.4$ rad) and radio (dashed lines) on-axis and off-axis. Off-axis light curves start lower and peak later than their on-axis counterparts. The simulation based curves are drawn as wide dotted lines. The bottom four plots show the relative differences between box and simulation for all four combinations, according to $(F_{box} - F_{sim})/F_{sim}$.}
 \label{compare_sim_figure}
\end{figure}

\begin{figure}
 \centering
 \includegraphics[width=1.0\columnwidth]{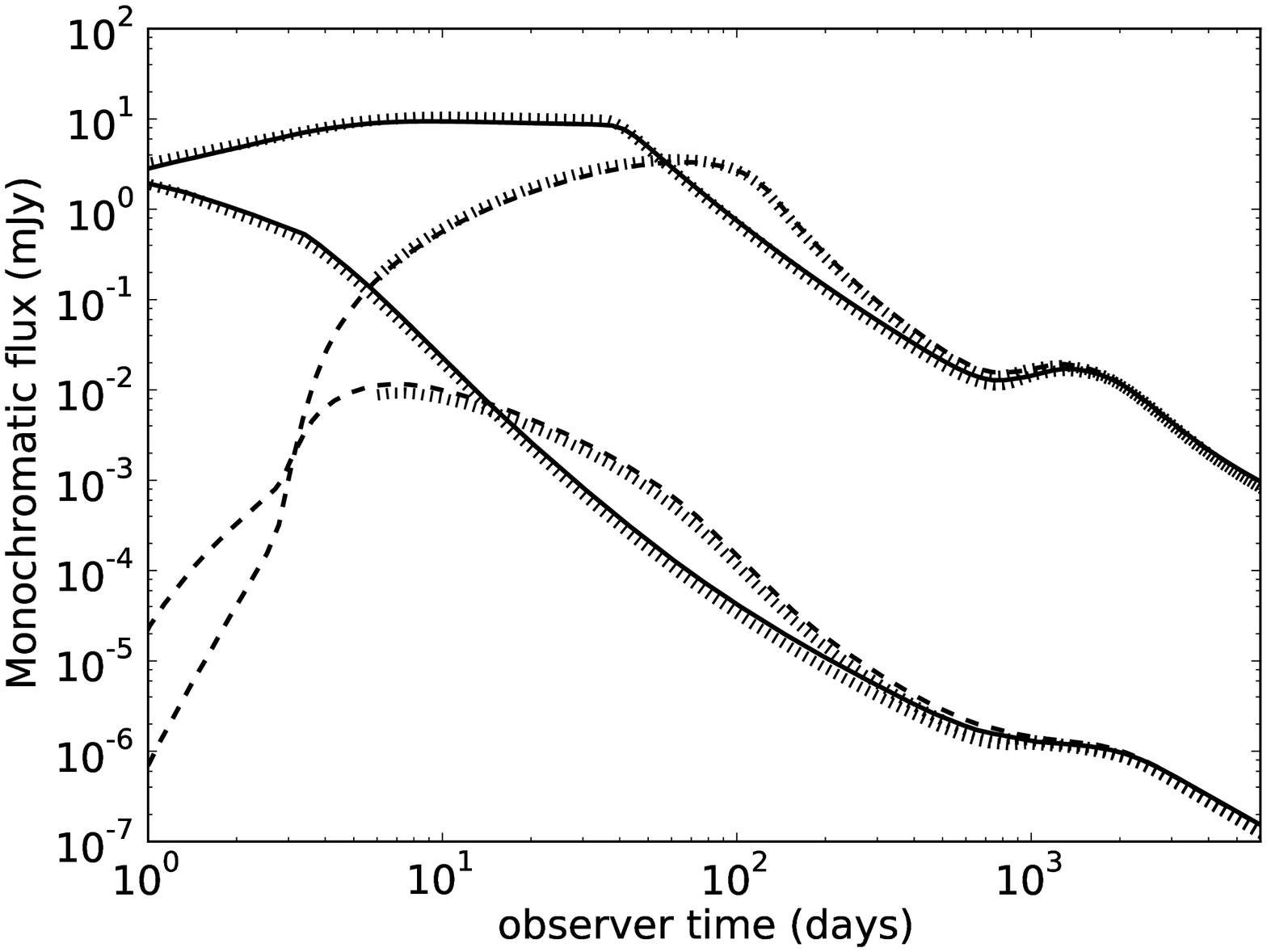}
 \caption{Direct comparison between box light curves and grid light curves from \citet[][see Figure 1 of that paper]{vanEerten2010c}. The simulation from that paper had $\theta_0 = 0.2$ rad, $E_{iso} = 10^{53}$ erg, $n_0 = 1$ cm$^{-3}$, $z = 0$, $d_L = 10^{28}$ cm, $p = 2.5$, $\epsilon_e = 0.1$, $\epsilon_B = 0.1$, $\xi_N = 1$, and the resulting light curves are indicated by thick dotted lines. The solid curves show the box results for $\theta_{obs} = 0$ rad at $10^9$ Hz (radio, top) and $10^{14}$ Hz (optical, bottom). The dashed curves show box results for $\theta_{obs} = 0.4$ rad, also at $10^9$ Hz (top) and $10^{14}$ Hz (bottom). Because the radiation was calculated by direct summation of the emitted power of all fluid elements in \cite{vanEerten2010c}, no synchrotron self-absorption was included in that work and for the purpose of comparison we have therefore disabled synchrotron self-absorption in the box curves for this plot as well.}
 \label{compare_offaxis_figure}
\end{figure}

\begin{figure}
 \centering
 \includegraphics[width=1.0\columnwidth]{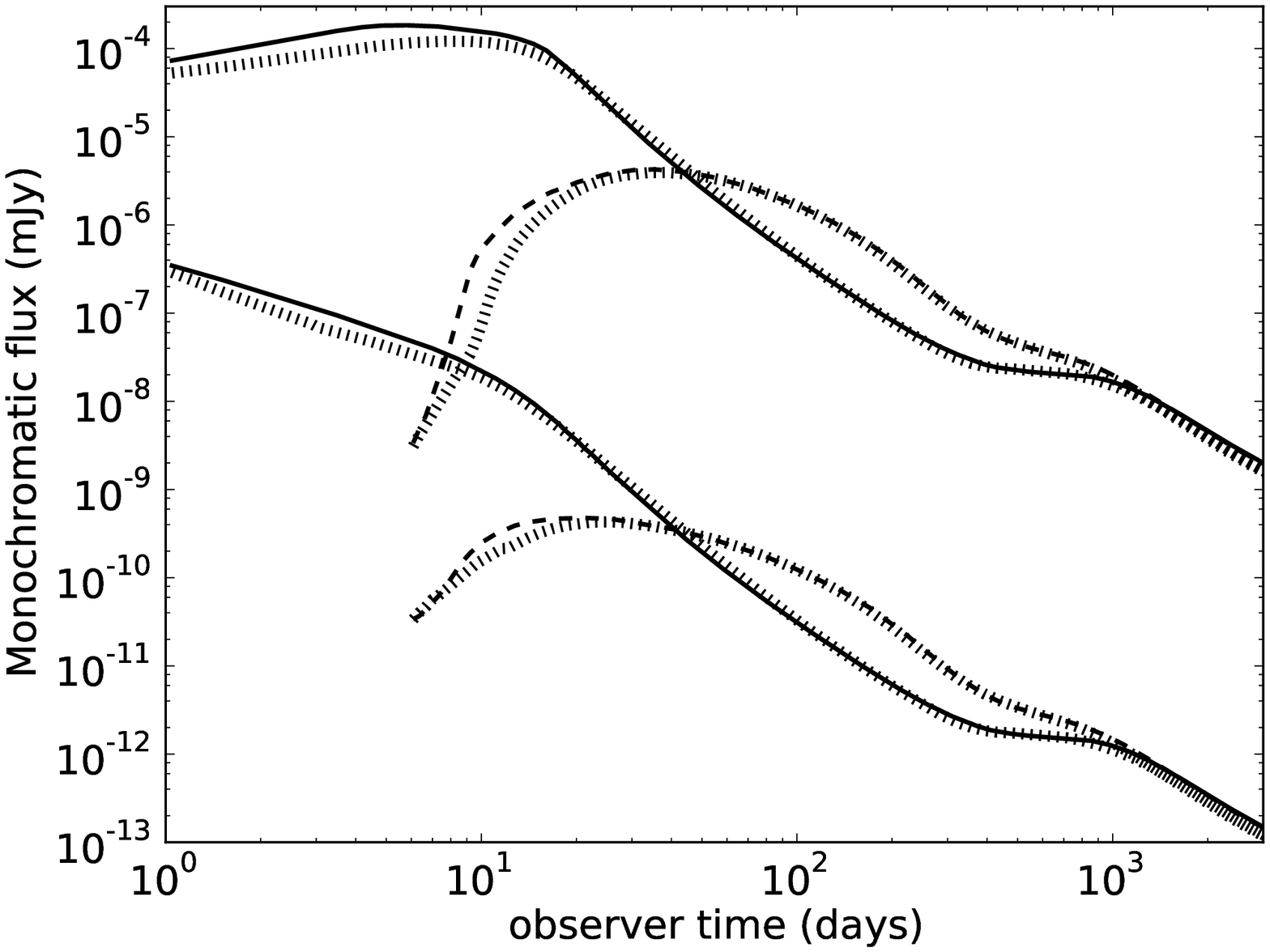}
 \caption{Direct comparison between box light curves and grid light curves from \citet[][see Figure 1 of that paper]{vanEerten2011sgrbs}. We use case B from that paper with $\theta_0 = 0.4$ rad, $E_{iso} = 1.25 \times 10^{49}$ erg, $n_0 = 10^{-3}$ cm$^{-3}$, $z = 0$, $d_L = 10^{28}$ cm, $p = 2.5$, $\epsilon_e = 0.1$, $\epsilon_B = 0.1$, $\xi_N = 1$, and the resulting light curves are indicated by thick dotted lines. The solid curves show the box results for $\theta_{obs} = 0$ rad at 1.43 GHz (radio, top) and $ 4.56 \times 10^{14}$ Hz (optical, bottom). The dashed curves show box results for $\theta_{obs} = 0.8$ rad, also at $1.43$ GHz (top) and $4.56 \times 10^{14}$ Hz (bottom). The emission coefficients in \cite{vanEerten2011sgrbs} were larger by a factor $3/2$ and in order to allow for a direct comparison, the fluxes from \cite{vanEerten2011sgrbs} have been divided by $3/2$.}
 \label{compare_sgrb_figure}
\end{figure}

We have performed various tests to check the resulting light curves from box based calculations. First of all, we have compared box based light curves to simulation based light curves for the simulation underlying the box summary. An example is shown in Figure \ref{compare_sim_figure}. At most, the difference between the two is on the order of a few percent (see bottom plots in figure). The exceptions are the early time light curves for observers far off-axis, when the box approach smoothens out numerical noise more than the direct simulation calculation does.

In Figures \ref{compare_offaxis_figure} and \ref{compare_sgrb_figure} we show a comparison to results of earlier studies. The close match between the box radio and optical light curves (solid and dashed lines for on and off-axis observer respectively) and the grid based counterparts (thick dotted lines) in Figure \ref{compare_offaxis_figure} is remarkable given the differences in the methods by which they were obtained. The grid based light curves are drawn from \cite{vanEerten2010c} and are therefore based on a simulation with different refinement and resolution settings and strongly differing isotropic energy compared to the unscaled simulations of the current work. In addition to that, the earlier curves have been calculated with a completely different radiation algorithm, where a summation was done over the emitted power of each fluid cell (which therefore excludes the possibility of synchrotron self-absorption), rather than by employing the linear radiative transfer method used in the current work. On average (over the logarithmically spaced data points) the difference between the box light curves and the light curves from the earlier study shown in the figure is a factor 1.15, with the biggest difference (a factor 1.31) occurring for the optical light curves (both on and off-axis) around 400 days. The fact that the off-axis light curves at some point cross the on-axis light curves and temporarily show higher flux levels is a result of relativistic beaming: at its most extreme it leads to the prediction of orphan afterglows, where the on-axis light curve remains effectively invisible relative to the off-axis light curve for observers at very high angles.

In Figure \ref{compare_sgrb_figure} we show a comparison to light curves from \cite{vanEerten2011sgrbs}. These were obtained from simulations with lower resolution compared to the current simulations, which started from $\gamma_b = 10$ rather than $\gamma_b = 25$. This accounts for the early time differences up to $\sim20$ days. Early time differences aside, the average difference (over the logarithmically spaced data points) is a factor 1.09, with the biggest difference (a factor 1.23) occurring at late times for the two off-axis light curves.

\begin{figure}
 \centering
 \includegraphics[width=1.0\columnwidth]{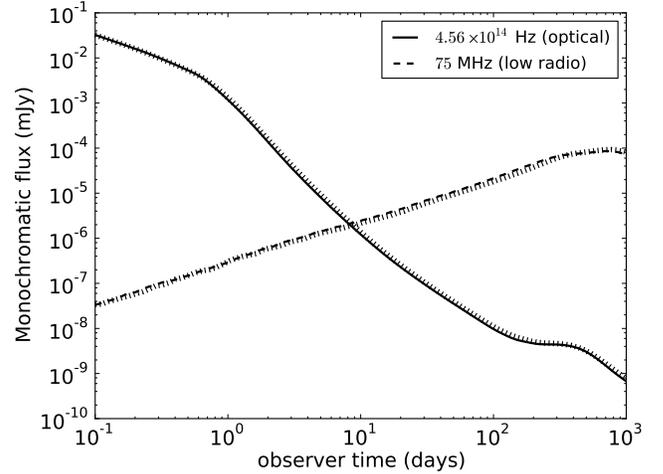}
 \caption{Direct comparison between box-based light curves with $\theta_0 = 0.2$ rad, where no interpolation with respect to jet opening angle has been applied (thick dotted curves, using only data from simulation 8 in Table \ref{sim_table}) and box-based light curves at $\theta_0 = 0.2$ rad created by interpolation from $\theta_0 = 0.175$ and $\theta_0 = 0.225$ rad (solid and dashed curves, using data from simulations 7 and 9).}
 \label{compare_int_figure}
\end{figure}

Finally, in Figure \ref{compare_int_figure} we show a comparison between optical and radio light curves based on a single opening angle simulation (0.2 rad) and light curves for the same opening angle reconstructed from interpolation between simulations with $\theta_0 = 0.175$ rad and $\theta_0 = 0.225$ rad. The figure shows that interpolated light curves generally match the light curves calculated from $\theta_0 = 0.2$ rad. In this particular example the greatest discrepancy occurs between the low radio curves around 160 days, where the interpolated curve flux is briefly larger by a factor 1.2. Note however that the figure represents an extreme scenario that does not occur in practice: by artificially removing $\theta_0 = 0.2$ for the purpose of testing the interpolation, we have tested interpolation across $\Delta \theta_0 = 0.025$ rad, whereas in practice the largest possible  difference $\delta \theta_0 = 0.0125$ rad. We conclude that the range of jet opening angles between $\theta_0 = 0.045$ and $\theta_0 = 0.5$ rad is adequately represented by our sample of simulations plus interpolation.

The numerical errors between different simulations and box based light curves given above should be compared to the difference between simulation/box based light curves on the one hand and analytically calculated light curves on the other hand. Although the latter light curves have no numerical noise or resolution errors by definition, they have systematic errors due to the simplifications in the underlying assumptions for dynamics and radiating region that are far larger than the numerical noise in the former. In \cite{vanEerten2010c} simulation results are compared to different analytical models and differences up to an order of magnitude in flux level and in time (for specific features such as the off-axis moment of peak flux) are seen, especially in the transrelativistic phase. 

\subsection{Fitting methods}
\label{fit_methods_section}

The method to quickly generate the observed flux at an arbitrary observer frequency and time described in the preceding subsections allows for iterative fitting of the simulation based afterglow model of a decelerating and spreading relativistic jet to broadband data. This model has at most 8 fit parameters: $E_{iso}$, $n_0$, $\theta_0$, $\theta_{obs}$ (observer angle), $\epsilon_e$, $\epsilon_B$, $\xi_N$, $p$. Observer luminosity distance $d_L$ and redshift $z$ are assumed to have been determined separately. Not all fit parameters need to be included in the fit and any parameter can be fixed to a specific value. For example, $\xi_N = 1$ and $\theta_{obs} = 0$ rad are commonly used.

The fit code takes as input the full set of broadband data points, all expressed in mJy. We use the downhill simplex method \citep{Nelder1965} combined with simulated annealing to minimize $\chi^2$ \citep{Kirkpatrick1983, Press1986}. We also use the suggestion from \cite{Nelder1965} to set the result for trial parameters outside of a specified parameter domain (e.g. $\theta_{obs} < 0$) equal to a very large number, which has the effect that the trial will be discarded before the downhill simplex iteration has completed.

The code is written in parallel. The broadband data points are distributed over the different computer cores and each core calculates the box-based flux counterpart for the data points it gets assigned, for each iteration of the fit parameters. Although the code can also be run on a single core, in practice the size of broadband data sets implies that even if the calculation of a single datapoint takes mere seconds, the total amount of calculation time required for the entire data set can become substantial. This is relevant especially in the case of an iterative fit that requires thousands of iterations (although strongly dependent on computer hardware, number of data points and numerical accuracy, the procedure can then still take days to complete).

In order to obtain a measure of the error on the fit variables, a Monte Carlo (MC) procedure is followed where the initial fluxes of the data set are randomly perturbed with an amplitude based on their error bars, and with a random number drawn from a Gaussian distribution, and then the fit is redone. This procedure is repeated a large number of times, i.e. $10,000$. We take the lowest 68.3\% of the resulting $\chi^2$ values and the extremes for the fit parameters within this subset determine their $1\sigma$ uncertainties.

Although the code allows for an MC calculation where the full box calculation is done each time a new model flux is required, the amount of time needed for the full MC run can become prohibitive. In order to circumvent the need for 10,000 data fits (consisting of thousands of flux calculations per data point each), the code offers an alternative approach for estimating the fit variable errors by calculating the light curves for fit variables other than the best fit from a series expansion in terms of the fit variables around the best fit. This means that instead of a complete flux calculation, at first the best fit result is calculated in detail. In addition to this the partial derivatives of the flux with respect to the different fit parameters are calculated. From the base light curve and the derivatives it is now possible to estimate the light curves at slightly differing values of the fit parameters. These are the values that will in practice be probed when $\chi^2$ is minimized for a perturbed data set.

Rather than the derivative $\partial F / \partial n_0$ (or any fit parameter other than $p, \theta_{obs}, \theta_0$), we use $\partial \log F / \partial \log n_0$ for this approach. The reason is that we know from analytical modeling that the flux in each spectral regime scales according to $F \propto E_{iso}^{\alpha_0} n_0^{\alpha_1} \epsilon_e^{\alpha_4} \epsilon_B^{\alpha_5} \xi_N^{\alpha_6}$ etc., where the coefficients $\alpha_i$ are either constant or linearly dependent on $p$ (see e.g. \citealt{Granot2001}). The method of calculating $\log F$ (rather than $F$) from some base value plus partial derivatives is therefore accurate beyond a mere first order approximation (in case of fixed $p$). Otherwise, the accuracy of the series expansion approach is set by the deviation of a given set of trial values from the best fit values. The maximum for this deviation is ultimately determined by the error on the data points. We did not use the logarithm of the angles because the observer angle can be equal to zero.

\section{GRB 990510: a case study}
\label{GRB990510_section}

The ``box''-based tool for GRB afterglow modeling presented in this paper can be applied to any broadband data set. Good spectral and temporal coverage is necessary to accurately determine all of the macro- and microphysical parameters. The broadband spectrum should span radio to X-ray frequencies to encompass all three characteristic frequencies of the synchrotron spectrum, and both early and late-time data are needed to determine for instance the opening angle of the jet and the observer angle. We note that the tool allows for fitting of limited data sets by fixing some of the fit parameters.

To demonstrate the capabilities of our tool, we have selected the afterglow of GRB~990510. There are light curves available for this source in X-rays, in various optical bands, and at two radio frequencies. Historically, GRB~990510 was the first strong case for an achromatic jet break, and the broadband afterglow has been modeled by several authors, all using analytical expressions. Here we show the results of our modeling with RHD simulations. 

The modeling tool requires fluxes in mJy at all frequencies, which means that some conversions have to be applied to the optical and X-ray data. The radio data at 4.8 and 8.7~GHz were taken directly from \citet{Harrison1999}. The X-ray count rates from \citet{Kuulkers2000} were converted to mJy using the conversion factors given in that paper. In our modeling we have used the V-, R- and I-band data from \citet{Harrison1999,Israel1999,Stanek1999,Pietrzynski1999,Bloom1999,Beuermann1999}. We have corrected the optical magnitudes for Galactic extinction with $E(B-V)=0.20$ \citep{Schlegel1998} before converting the magnitudes into fluxes. \citet{Starling2007} have shown that the host galaxy extinction is negligible for GRB~990510 based on modeling of the X-ray to optical spectrum.

\subsection{Fit results}

\begin{table*}
\centering
\caption{Best-fit model parameters and 1 $\sigma$ errors. The column labeled PK02 lists the fit results from \cite{Panaitescu2002}, translated from their units and 90\% confidence intervals. For context we also include jet energies $E_j$ results.}
\begin{tabular}{|l|llll|}
\hline
 var. & PK02 & on-axis, fixed $\xi_N$ & on-axis & off-axis \\
\hline
 $\theta_0$ (rad)    & $5.4^{+0.1}_{-0.6}  \times 10^{-2}$       &     $7.5^{+0.2}_{-0.4} \times 10^{-2}$          & $9.46_{+0.03}^{-0.33} \times 10^{-2}$            & $4.82^{+0.32}_{-0.04} \times 10^{-2}$ \\
 $E_{iso}$ (erg)     & $9.47^{+37}_{-2.27} \times 10^{53}$  &  $1.8^{+0.3}_{-0.1} \times 10^{53}$  &  $1.04^{+0.16}_{-0.02} \times 10^{53}$ & $4.388^{+0.003}_{-0.605} \times 10^{54}$ \\
 $n_0$ (cm$^{-3}$)         & $2.9^{+0.7}_{-0.9} \times 10^{-1} $       & $3.0_{-1.2}^{+0.4} \times 10^{-2} $  & $1.15^{+0.03}_{-0.19}$             & $1.115^{+0.258}_{-0.006} \times 10^{-1}$ \\
 $\theta_{obs} (rad)$ & 0 (fixed)                                & 0 (fixed)                                & 0 (fixed)                             & $1.6^{+0.1}_{-0.1} \times 10^{-2}$ \\
 $p$            & $1.83^{+0.11}_{-0.006}$        &      $2.28^{+0.06}_{-0.01}$      & $2.053_{+0.007}^{-0.006}$                & $2.089^{+0.013}_{-0.001}$ \\
 $\epsilon_B$   & $5.2^{+26}_{-2.9} \times 10^{-3}$  &  $4.6^{+0.8}_{-0.8} \times 10^{-3}$  &  $2.04^{+0.04}_{-0.30} \times 10^{-3}$ & $1.36^{+0.19}_{-0.03} \times 10^{-3}$ \\
 $\epsilon_e$   & $2.5^{+1.9}_{-0.4} \times 10^{-2}$ & $3.73_{+0.07}^{-0.68} \times 10^{-1}$  & $6.8^{+0.6}_{-0.1} \times 10^{-1}$ & $1.17^{+0.02}_{-0.12} \times 10^{-2}$ \\
 $\xi_N$        & 1 (fixed)    & 1 (fixed)                      & $5.4^{+0.6}_{-0.6} \times 10^{-1}$               & $5.7^{+1.0}_{-1.7} \times 10^{-2}$ \\
\hline
$\chi^2_r$  &  ...  & $6.389$ & $5.389$ & $3.235$ \\
$E_j$      & $1.4^{+3.1}_{-0.3} \times 10^{50}$  &  $5.0^{+1.2}_{-0.8} \times 10^{50}$  & $4.63^{+0.05}_{-0.10} \times 10^{50}$ & $5.1^{+0.7}_{-0.8} \times 10^{51}$ \\
\hline
\end{tabular}
\label{fit_values_table} 
\end{table*}

\begin{figure}
 \centering
 \includegraphics[width=1.0\columnwidth]{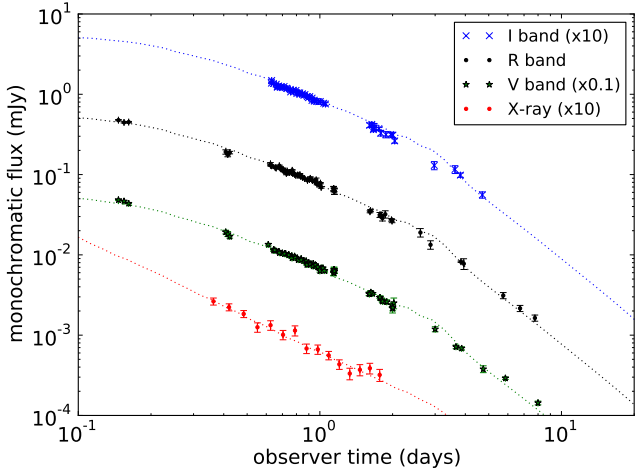}
 \includegraphics[width=1.0\columnwidth]{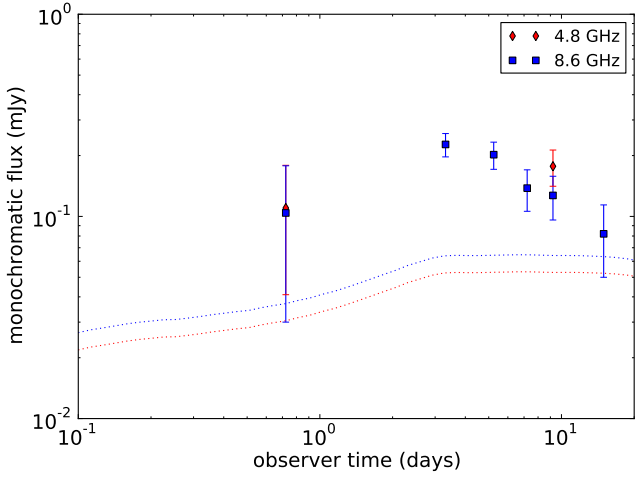}
 \caption{Fit results for the GRB 990510 on-axis fit ($\theta_{obs} = 0$) with fixed $\xi_N=1$. The reduced $\chi^2$ for 205 data points and 6 fit parameters is 6.389. For clarity of presentation, some fluxes have been multiplied by the indicated factors.}
 \label{lc12noksiplot_figure}
\end{figure}

\begin{figure}
 \centering
 \includegraphics[width=1.0\columnwidth]{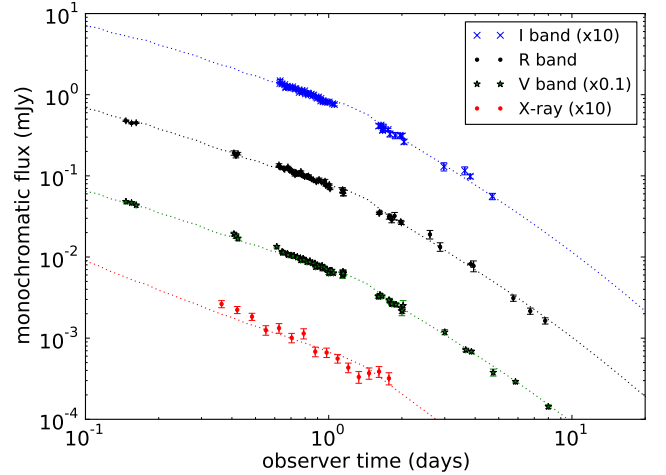}
 \includegraphics[width=1.0\columnwidth]{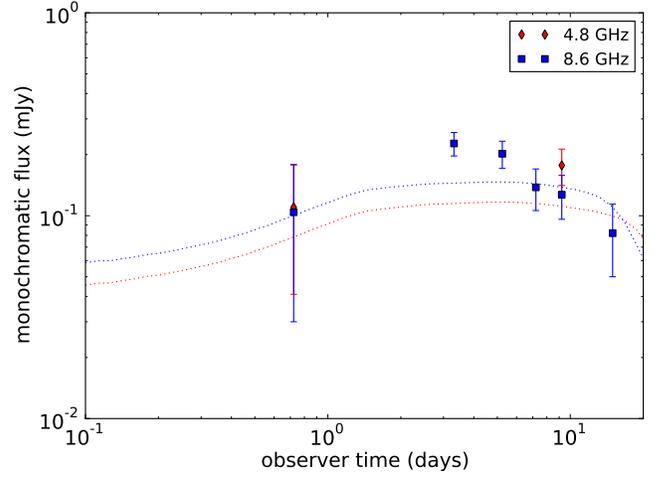}
 \caption{Fit results for GRB 990510 broadband on-axis fit ($\theta_{obs} = 0$). The reduced $\chi^2$ for 205 datapoints and 7 fit parameters is 5.389. For clarity of presentation, some fluxes have been multiplied by the indicated factors.}
 \label{lc12plot_figure}
\end{figure}

\begin{figure}
 \centering
 \includegraphics[width=1.0\columnwidth]{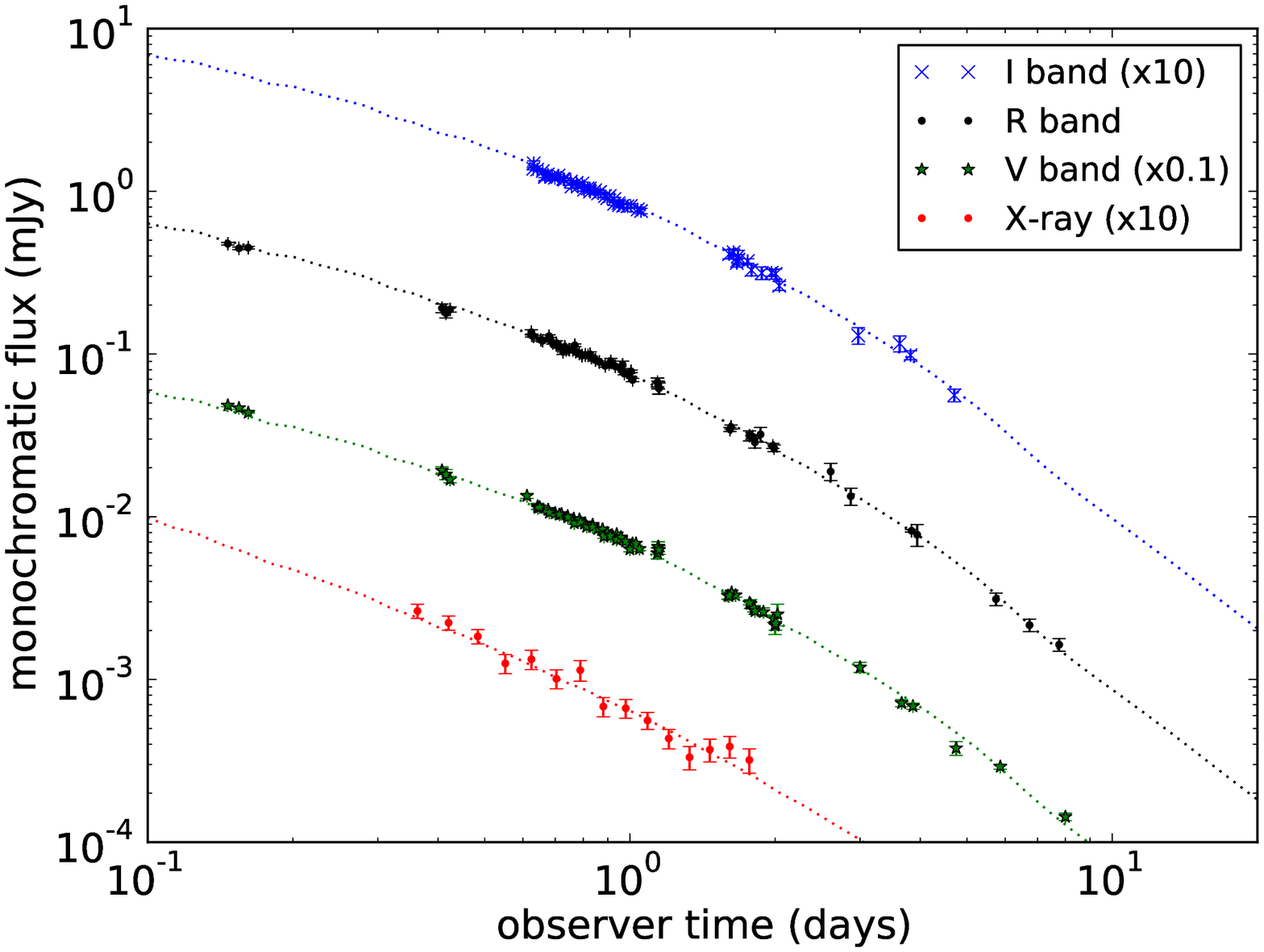}
 \includegraphics[width=1.0\columnwidth]{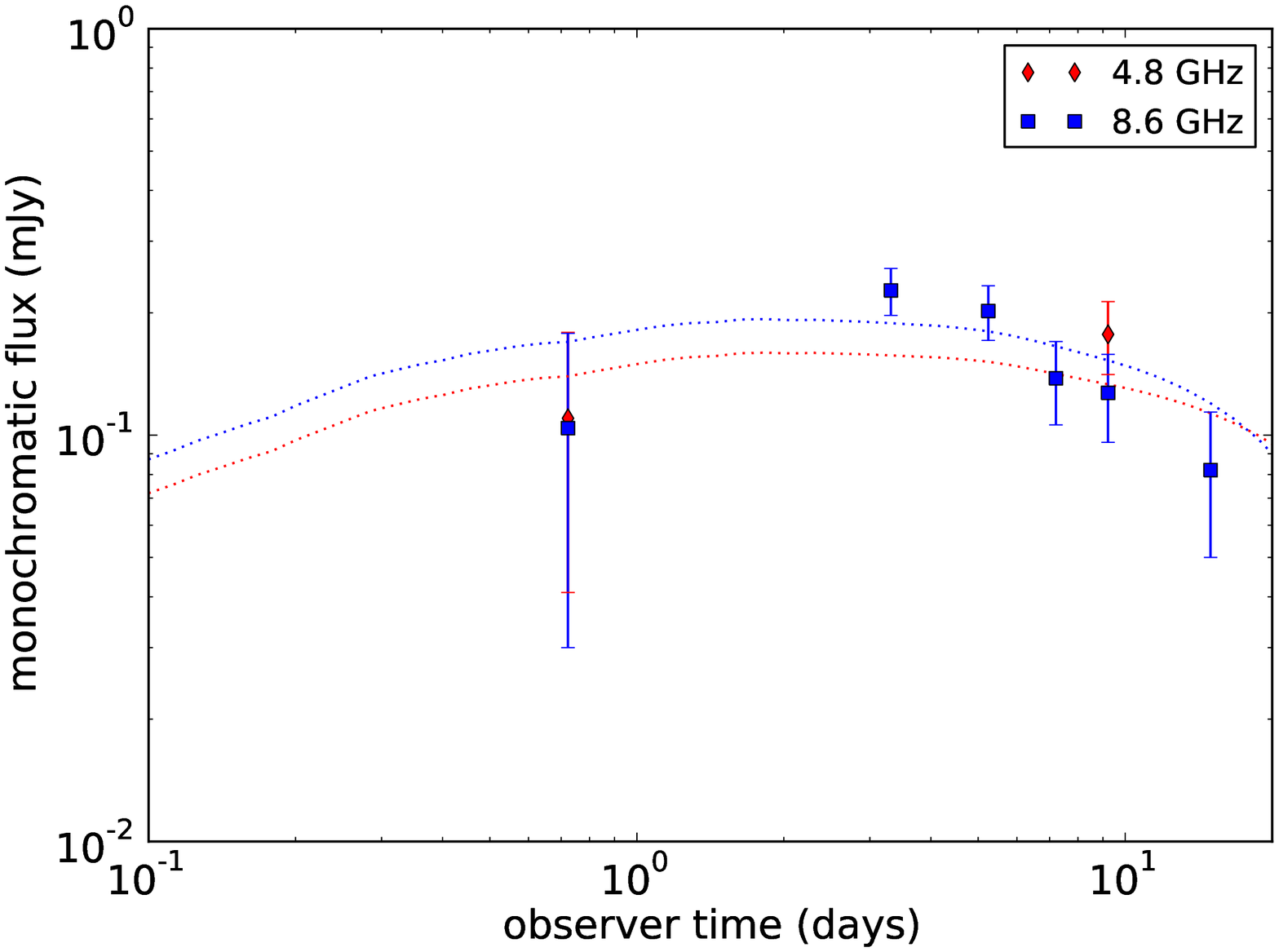}
 \caption{Fit results for GRB 990510 broadband off-axis fit, with the observer angle included as a fit parameter. The reduced $\chi^2$ for 205 datapoints and 8 fit parameters is 3.235. For clarity of presentation, some fluxes have been multiplied by the indicated factors.}
 \label{lc12offplot_figure}
\end{figure}

We have performed three different fits to the 205 data points of the GRB~990510 afterglow, for which we show the resulting fit parameters in Table \ref{fit_values_table} and the accompanying best fit light curves in Figures \ref{lc12noksiplot_figure}$-$\ref{lc12offplot_figure}. Figure \ref{lc12noksiplot_figure} shows the fit results for a fit with fixed $\theta_{obs}=0$ and $\xi_N=1$, which can be directly compared to the broadband fits performed by \citet{Panaitescu2002} who use analytical expressions. Figure  \ref{lc12plot_figure} shows the best fit light curves for a fit with $\xi_N$ as a free parameter and fixed $\theta_{obs}=0$, and Figure \ref{lc12offplot_figure} for a fit without any fixed parameters. 

Table \ref{fit_values_table} shows a clear spread in best fit parameters for the three fits we performed and also some differences with the results from \citet{Panaitescu2002}. These differences can be mostly attributed to the fact that the synchrotron self-absorption frequency $\nu_a$ is not very well defined by this particular data set. The coverage at radio frequencies is fairly sparse compared to the optical and X-ray bands, and the flux uncertainties in the radio are also larger than at higher frequencies. For all three of our fits $\nu_a$ has a value around $10^9$~Hz at 1~day after the burst, while the peak frequency $\nu_m\sim10^{13}$~Hz at that time. The good coverage in the optical and X-ray bands enables an accurate determination of the cooling frequency $\nu_c$, which is situated just above the optical bands, and the value of $p$. 
In contrast to the results from \cite{Panaitescu2002}, we find $p > 2$ for all three our fits rather than $1.8$. Although our $p > 2$ lies within the error bar of their work, the converse is not true, which confirms $p > 2$, and that there is thus no need to include an additional high energy cut-off on the relativistic particle distribution. The value of $p$ has also been determined by several other authors based on optical and X-ray light curve slopes \citep[e.g.][]{Harrison1999,Kuulkers2000,Panaitescu2001a} or a combination of light curves and optical to X-ray spectra \citep{Starling2008}. In those studies the value for $p$ falls in the range $2.1-2.2$, consistent with the $p$ value we have obtained in our fit without any fixed parameters. 

It is clear from Table \ref{fit_values_table} that adding extra parameters introduces quite a strong variation in some of the parameters. In particular, the energy and circumburst density change from the on-axis to the off-axis fit with more than an order of magnitude, while in the off-axis fit the opening angle becomes a factor of two smaller by introducing a non-zero observer angle. This shows the importance including the observer angle in broadband modeling of GRB afterglows, although this does require well-sampled light curves across the whole spectrum. More detailed studies of well-sampled broadband afterglows will be presented in a future paper.

\section{Discussion}
\label{discussion_section}

In this paper we present a method to directly fit light curves based on 2D hydrodynamical jet simulations to broadband afterglow data. This provides a clear improvement over fits based on analytical models, which are not able to take complex features of the jet dynamics into account, such as the radial fluid profile, slow deceleration from ultra-relativistic to non-relativistic outflow, the sideways spreading, and resulting inhomogeneity along the shock front. The iterative fit procedure is possible because $(i)$ we have shown that the jet evolution is scale invariant with respect to explosion energy and circumburst medium density, and $(ii)$ the results from high-resolution parallel RHD simulations can be summarized in a very compact form. The compressed version, a `box' summary, of the simulation `grid' data is possible once the blast wave lateral extent, radius and radial width are known from the data at each emission time. The predicted flux is calculated for each data point for a given set of explosion and radiation parameters, and because the code takes into account both electron cooling and synchrotron self-absorption, it is applicable to the entire broadband afterglow spectrum. 

In order to set up the boxes, a total of 19 RHD simulations were performed in 2D, with initial opening angles varying between 0.045 and 0.5 radians. Light curves for intermediate opening angles between different tabulated simulation results are obtained via interpolation at the fluid level. The simulations are run for a long time in order to ensure that late time non-relativistic features such as the rise of the counterjet are covered as well. At very early times the outflow conforms to the self-similar Blandford-McKee solution and this is used directly to calculate radiation from emission times before the starting point of the simulations. A comparison of the evolution of the 19 simulations reveals that, although hard to predict analytically, the evolution from strongly collimated BM jet to semi-spherical Sedov-Taylor blast wave is smooth. For small opening angles, the intermediate stage between the two asymptotes is more pronounced.

We present a number of tests for the resolution of box-based light curves by direct comparison to the simulations underlying the box summaries and to earlier work. The earlier light curves have been generated directly from simulations using different methods: by applying linear radiative transfer, and by summation of the emitted power (in the optically thin case). The differences between box based and simulation based light curves are found to be a few percent at most, and the box summaries are found to correctly capture the shock profile at the different stages of blast wave evolution.

We have used GRB~990510 as a case study to demonstrate our method, because it has been observed over a wide range of frequencies. A simultaneous fit has been performed to data at two radio frequencies, three optical bands and in X-rays. There are some substantial differences between our best fit values and earlier fit values found by \cite{Panaitescu2002}, but also between the fits with various parameters fixed that we have performed. Most strikingly, including a non-zero observing angle in our modeling changes has a large impact on the obtained values for the blast wave energy and circumburst density. Furthermore, in contrast with \cite{Panaitescu2002} we find a value for the electron energy distribution index $p>2$ and we thus do not need to include a high energy cut-off on the relativistic particle distribution.

The more accurate fits possible by the method of this paper help to further constrain the physics that shape the gamma-ray burst afterglows. Explosion energy and circumburst density provide important clues to the nature of the progenitor star and burst environment. In addition, the type of fits to data sets covering a long time span that the method makes possible, where the complexities of the intermediate stage dynamics and the shape of the jet break are fully included, are necessary to establish a baseline for studies of the effect of more detailed models of the microphysics. The evolution of microphysical parameters, such as $\epsilon_B$, is discussed in various papers \citep{vanEerten2010, Panaitescu2006, Filgas2011}.

Even without utilizing the possibility of fitting broadband data directly to simulation results, the fact that the box summaries cover not just a large number of simulations directly, but through scaling and interpolation, exploring the complete parameter space of impulsive jets in an ISM environment (with the inclusion of stellar wind environments being straightforward) should prove useful. It will allow us to test radiation mechanisms of physical interest other than pure synchrotron radiation in a realistic context. As long as there is no significant feedback on the dynamics of the outflow or dynamically relevant magnetic field involved, any generalization is possible even if it includes scattering. Examples of interest for different radiative processes are given by \cite{Giannios2009, Petropoulou2009}.

The source code of the broadband fit program described in this paper is publicly available at \url{http://cosmo.nyu.edu/afterglowlibrary}. The code has different settings: not only can it be used to fit box based light curves to a data set and establish the uncertainties in the best fit parameters, it can also be used to generate light curves and spectra directly for arbitrary frequencies, observer angles, observer times, and explosion and radiation parameters. This could be helpful for directly exploring how the different regions of the parameter space determine the shape of the afterglow, and for quickly creating light curves that can be expected or looked for in surveys (see e.g. \citealt{Nakar2011,Roberts2011,Metzger2011}).

A number of practical improvements can be made to the code. In the case of very large data sets, even for the parallel version, where the datapoints to be calculated are evenly distributed on the cores, the total calculation time can become unwieldy. A remedy to this would be to no longer recalculate the flux independently for each data point but to calculate the flux at a fixed (large) number of values for observer time and frequency, chosen such to evenly cover the available data. The flux at the exact datapoint values can then be determined by interpolation between these values. Because the light curves are smooth, this will not impact the accuracy of the fit. For very long data sets (some GRBs have now been observed for several years, e.g. GRB\,030329,  \citealt{Frail2000,VanDerHorst2008}), more long term simulation data can be added to the boxes. Boxes can also be generated for a stellar wind environment, since the same scale invariances apply. These applications and improvements will be presented in future work.

%\clearpage

\acknowledgments
%\section{Acknowledgements}
This research was supported in part by NASA through grant NNX10AF62G issued through the Astrophysics Theory Program and by the NSF through grant AST-1009863. Resources supporting this work were provided by the NASA High-End Computing (HEC) Program through the NASA Advanced Supercomputing (NAS) Division at Ames Research Center. The software used in this work was in part developed by the DOE-supported ASCI/Alliance Center for Astrophysical Thermonuclear Flashes at the University of Chicago. We would like to thank Erik Kuulkers for providing the X-ray fluxes. HJvE acknowledges hospitality from NASA/MSFC in Huntsville, where part of this research was conducted, and AJvdH likewise acknowledges hospitality from New York University. Finally, HJvE \& AJvdH gratefully acknowledge Ralph Wijers whose PhD supervision and research program at the University of Amsterdam have laid the foundation for their contributions to the current study.\\

\appendix

\section{emission and absorption coefficients}
\label{coefficients_section}
The expressions for the emission and absorption coefficient are drawn from \cite{Granot1999}, based in turn on \citealt{Sari1998}, and for completeness we provide below the exact forms that we have implemented in our radiation code\footnote{The terminology used in \cite{vanEerten2010c} is slightly off, where the term \emph{emissivity} has been used to refer to a quantity that should have been labeled \emph{emission coefficient}, had it not been for the factor $4 \pi$ that was left explicit as well. The terminology has been corrected for the current paper and now matches \cite{Rybicki1979}. The coefficient $j_S$ in \cite{vanEerten2011sgrbs} was larger by a factor $3/2$.}. The code solves the radiative transfer equation using
\begin{equation}
\Delta I_\nu = (j_\nu - \alpha_\nu I_\nu) c \Delta t,
\end{equation}
where $\Delta t$ the lab frame time difference between two snapshots, $j_\nu$ the emission coefficient and $\alpha_\nu$ the absorption coefficient. The code subsequently calculates the flux by integrating over the area $A$ of the emission plane, according to
\begin{equation}
F_\nu = \frac{1+z}{d_L^2} \int \dev{A} I_\nu.
\end{equation}
Here $d_L$ is the luminosity distance and $z$ the redshift. Separating the coefficients into frequency dependent and non-frequency dependent components, $j_\nu = j_S \times j_f$ and $\alpha_\nu = \alpha_S \times \alpha_f$, we use for the non-frequency dependent components:
\begin{equation}
 j_S = 9.6323 \frac{p-1}{3p-1}\frac{\sqrt{3} q_e^3}{8 \pi m_e c^2} \frac{\xi_N n' B'}{\gamma^2 (1 - \beta \mu)^2}, \qquad \alpha_S = \frac{\sqrt{3} q_e^3 (p-1)(p+2)}{16 \pi m_e^2 c^2} \xi_N n' B' \gamma (1 - \beta \mu).
\end{equation}
Here $q_e$ is the electron charge, $m_e$ the electron mass, $n'$ the comoving number density, $B'$ the comoving magnetic field strength, $\beta$ the fluid flow velocity as fraction of $c$, and $\mu$ the angle between the outflow and the observer direction. The frequency dependent part $j_f$ is given by
\begin{equation}
 j_f = \left\{
\begin{array}{rl}
 ( \nu' / \nu'_m )^{1/3} & \text{if } \nu' < \nu'_m < \nu'_c, \\
 ( \nu' / \nu'_m )^{(1-p)/2} & \text{if } \nu'_m < \nu' < \nu'_c, \\
 ( \nu'_c / \nu'_m)^{(1-p)/2} (\nu' / \nu'_c)^{-p/2} & \text{if } \nu'_m < \nu'_c < \nu',
\end{array} \right.
\end{equation}
when $\nu'_m < \nu'_c$ and by
\begin{equation}
 j_f = \left\{
\begin{array}{rl}
 ( \nu' / \nu'_c )^{1/3} & \text{if } \nu' < \nu'_c < \nu'_m, \\
 ( \nu' / \nu'_c )^{-1/2} & \text{if } \nu'_c < \nu' < \nu'_m, \\
 ( \nu'_m / \nu'_c )^{-1/2} ( \nu' / \nu'_m )^{-p/2} & \text{if } \nu'_c < \nu'_m < \nu',
\end{array} \right.
\end{equation}
 otherwise. Here $\nu'$ denotes the comoving observer frequency. The synchrotron frequency $\nu'_m$ is given by
\begin{equation}
 \nu'_m = \frac{3 q_e}{4 \pi m_e c} (\gamma'_m)^2 B', \qquad \gamma'_m = \left( \frac{p-2}{p-1} \right) \left( \frac{\epsilon_e e'}{\xi_N n' m_e c^2} \right).
\end{equation}
The cooling break frequency $\nu'_c$ is estimated using a global cooling time, leading to
\begin{equation}
 \nu'_c = \frac{3 q_e}{4 \pi m_e c} (\gamma'_c)^2 B', \qquad \gamma'_c = \frac{6 \pi m_e c \gamma}{\sigma_T (B')^2 t_c}, \qquad t_c \equiv t.
\end{equation}
Note that both $\gamma'_m$ and $\gamma'_c$ are comoving with the fluid, while $\gamma_c$ in \cite{Sari1998} is in the rest frame.

For the self-absorption coefficient we ignore the effects of cooling, because in practice the self-absorption break frequency $\nu'_a \ll \nu'_c$ and because otherwise the limit of accuracy is set in practice by the use of a global cooling time rather than by any additional level of detail in the calculation of the absorption coefficient. We have
\begin{equation}
 \alpha_f = \frac{1}{\gamma'_m (\nu')^2} \left\{
\begin{array}{rl}
 ( \nu' / \nu'_m )^{1/3} & \text{if } \nu' < \nu'_m, \\
 ( \nu' / \nu'_m )^{-p/2} & \text{if } \nu'_m < \nu'. \\
\end{array} \right.
\end{equation}

\section{Scale-free fluid equations}
\label{scale_invariance_appendix}

In our study we have solved the fluid equations using arbitrary values for explosion energy and density. For reference and in order to explictly demonstrate the complete scale-invariance of the simulations we provide the fluid equations in terms of dimensionless parameters $A$, $B$, $\theta$ below.

The special relativistic fluid dynamics equations in spherical coordinates, assuming symmetry in angle $\phi$ in the $x-y$ plane around the jet axis, are as follows:
\begin{eqnarray}
 \frac{\partial}{ct} \gamma \rho + (\frac{\partial}{\partial r} + \frac{2}{r}) \gamma \rho \beta_r + \frac{1}{r \sin \theta} \frac{\partial}{\partial \theta} \gamma \rho \beta_\theta \sin \theta& = & 0, \nonumber \\
 \frac{\partial}{\partial ct} (h \gamma^2 - p - \rho \gamma c^2) + (\frac{\partial}{\partial r} + \frac{2}{r} ) (h \gamma^2 \beta_r - \rho \gamma c^2 \beta_r) + \frac{1}{r \sin \theta} \frac{\partial}{\partial \theta} [(h \gamma^2 \beta_\theta - \rho \gamma c^2 \beta_\theta) \sin \theta ] & = & 0, \nonumber \\
 \frac{\partial}{ct} h \gamma^2 \beta_r + (\frac{\partial}{\partial r} + \frac{2}{r}) (h \gamma^2 \beta_r^2 + p) + \frac{1}{r \sin \theta} \frac{\partial}{\partial \theta} h \gamma^2 \beta_r \beta_\theta & = & 0, \nonumber \\
\frac{\partial}{ct} h \gamma^2 \beta_\theta + (\frac{\partial}{\partial r} + \frac{2}{r}) h \gamma^2 \beta_\theta \beta_r + \frac{1}{r \sin \theta} \frac{\partial}{\partial \theta} (h \gamma^2 \beta_\theta^2 + p) & = & 0.
\end{eqnarray}
Here $\beta_r$ and $\beta_\theta$ are the fluid velocity components in the $r$ and $\theta$ direction respectively in units of $c$ and $h$ the relativistic enthalpy density including rest mass energy density. In terms of scale-free parameters $A$ and $B$, the partial derivatives in $r$ and $ct$ can be expressed as
\begin{eqnarray}
 \frac{\partial}{\partial ct} & = & - \frac{A}{ct} \frac{\partial}{\partial A} + \frac{2B}{ct} \frac{\partial}{\partial B}, \nonumber \\
 \frac{\partial}{\partial r} & = & \frac{1}{ct} \frac{\partial}{\partial A} - \frac{5B}{r} \frac{\partial}{\partial B}.
\end{eqnarray}
Combining these with the fluid equations above yields the scale-invariant forms
\begin{eqnarray}
 (-A \frac{\partial}{\partial A} + 2B \frac{\partial}{\partial B}) \gamma \frac{\rho}{\rho_0} + (\frac{\partial}{\partial A} - \frac{5 B}{A} \frac{\partial}{\partial B} + \frac{2}{A}) \gamma \frac{\rho}{\rho_0} \beta_r + \frac{1}{A \sin \theta} \frac{\partial}{\partial \theta} \gamma \frac{\rho}{\rho_0} \beta_\theta \sin \theta & = & 0, \nonumber \\
(-A \frac{\partial}{\partial A} + 2B \frac{\partial}{\partial B}) \frac{h \gamma^2 - p - \rho \gamma c^2}{\rho_0 c^2} + (\frac{\partial}{\partial A} - \frac{5 B}{A} \frac{\partial}{\partial B} + \frac{2}{A}) \frac{h \gamma^2 \beta_r - \rho \gamma c^2 \beta_r}{\rho_0 c^2} + \frac{1}{A \sin \theta} \frac{\partial}{\partial \theta} \frac{(h \gamma^2 \beta_\theta - \rho \gamma c^2 \beta_\theta) \sin \theta}{\rho_0 c^2} & = & 0, \nonumber \\
(-A \frac{\partial}{\partial A} + 2B \frac{\partial}{\partial B}) \frac{h \gamma^2 \beta_r}{\rho_0 c^2} + (\frac{\partial}{\partial A} - \frac{5 B}{A} \frac{\partial}{\partial B} + \frac{2}{A}) \frac{h \gamma^2 \beta_r^2 + p}{\rho_0 c^2} + \frac{1}{A \sin \theta} \frac{\partial}{\partial \theta} \frac{h \gamma^2 \beta_r \beta_\theta}{\rho_0 c^2} & = & 0, \nonumber \\
(-A \frac{\partial}{\partial A} + 2B \frac{\partial}{\partial B}) \frac{h \gamma^2 \beta_\theta}{\rho_0 c^2} + (\frac{\partial}{\partial A} - \frac{5 B}{A} \frac{\partial}{\partial B} + \frac{2}{A}) \frac{h \gamma^2 \beta_\theta \beta_r}{\rho_0 c^2} + \frac{1}{A \sin \theta} \frac{\partial}{\partial \theta} \frac{h \gamma^2 \beta_\theta^2 + p}{\rho_0 c^2} & = & 0.
\end{eqnarray}
Quantities such as $\gamma \rho / \rho_0$ etc. are dimensionless and therefore (scale-invariant) functions of the scale-invariant dimensionless parameters $A$, $B$ and $\theta$. For radial outflow and for limiting values of $A$, the fluid equations further reduce to self-similarity.
\bibliography{boxfit}

\end{document}